\RequirePackage{ifpdf}
\documentclass{PoS}
\pdfoutput = 1
\usepackage{cite}
\usepackage{amsmath}  
\usepackage{amssymb} 
\usepackage{dsfont}
 \usepackage{multirow}
\usepackage{psfrag,scalefnt}
\usepackage{paralist}

\usepackage{color}
\let\normalcolor\relax

\newcommand{\vcb}{|V_{cb}|}
\newcommand{\vtd}{|V_{td}|}
\newcommand{\vub}{|V_{ub}|}
\newcommand{\vts}{|V_{ts}|}

\def\epe{\varepsilon'/\varepsilon}
\newcommand{\tev}{\, {\rm TeV}}

\newcommand{\be}{\begin{equation}}
\newcommand{\ee}{\end{equation}}
\newcommand{\bea}{\begin{eqnarray}}
\newcommand{\eea}{\end{eqnarray}}

\newcommand{\ba}{\begin{array}}
\newcommand{\ea}{\end{array}}

\newcommand{\ord}{{\cal O}}

\def\kpn{K^+\rightarrow\pi^+\nu\bar\nu}

\def\klpn{K_{L}\rightarrow\pi^0\nu\bar\nu}

\title{Flavour Expedition to the Zeptouniverse}

\ShortTitle{Flavour Expedition to the Zeptouniverse}

\author{\speaker{Andrzej J. Buras}%
 \thanks{FLAVOUR(267104)-ERC-90} \\
TUM-IAS, Lichtenbergstr. 2a, D-85748 Garching, Germany \\
Technical University Munich, Physics Department, D-85748 Garching, Germany,\\
E-mail: \email{aburas@ph.tum.de}}

\abstract{After the completion of the Standard 
Model (SM) through the Higgs discovery in 2012 particle physicists are waiting for the discovery of new particles 
either directly with the help of the Large Hadron Collider (LHC) or indirectly 
through quantum fluctuations causing certain rare processes to occur at  different rates than predicted by the SM. While the later route is very challenging, 
requiring very precise theory and experiment, it allows in principle a much 
higher resolution of short distance scales than it is possible with the help 
of the LHC. In fact, in the coming flavour precision era, in which the 
accuracy of the measurements of rare processes and of the relevant lattice QCD calculations will be significantly increased, there is a good chance that we may get 
an insight into the scales as short as $10^{-21}$ m ({\it Zeptouniverse}) corresponding to energy scale of $200\tev$ or even shorter distance scales. We discuss the requirements that have to be met 
for such a flavour expedition to the Zeptouniverse to be successful.
In particular we emphasize the power of correlations between flavour observables in the search for New Physics (NP) and  identify 
a number of correlations that could allow to discover NP even if it would appear at the level of $20\%$ of the SM contributions.  The correlation between $\mathcal{B}(\kpn)$, $\overline{\mathcal{B}}(B_s\to\mu^+\mu^-)$  and $\gamma$ extracted from tree-level decays within the SM is one of them.
After presenting the simplest correlations in CMFV and $U(2)^3$ models 
we address the recent data on $B_{s,d}\to\mu^+\mu^-$ and the 
anomalies in $B_d\to K(K^*)\mu^+\mu^-$, including  breakdown of lepton 
flavour universality, in the context of $Z^\prime$-models  
 with quark flavour violating neutral couplings. A brief discussion of leptoquark models is also given. We emphasize the correlations of  $B_d\to K(K^*)\mu^+\mu^-$ with $B_d\to K(K^*)\nu\bar\nu$ 
that allow to distinguish between various explanations of the 
anomalies in question. Finally, we summarize the recent study of $K\to\pi\nu\bar\nu$ and 
$B_{d,s}\to\mu^+\mu^-$ decays which demonstrates that these decays play important roles in finding out what happens in the Zeptouniverse. 
}

\FullConference{Flavorful Ways to New Physics\\
                 28-31 October 2014\\
                 Freudenstadt - Lauterbad, Germany}

\begin{document}

\section{Overture}
In spite of tremendous efforts of experimentalists and theorists to find 
New Physics (NP) beyond the Standard Model (SM), no clear indications for NP beyond dark matter, neutrino masses and matter-antimatter asymmetry in the universe have been observed. Yet, the recent discovery of a Higgs-like particle and the overall agreement of the SM with the present data shows that our general approach of describing physics at very short distance scales with the help of exact (QED and QCD) and spontaneously broken (for weak interactions) gauge theories is correct.

As the SM on the theoretical side is not fully satisfactory and the three 
NP signals mentioned above are already present, we know that some new particles 
and new forces have to exist, hopefully within energy scales being directly explored by the LHC or not far above them.
The upgrade in the energy of the LHC, the upgrade of the LHCb, SuperKEKB and dedicated 
kaon physics  experiments at CERN and J-PARC, as well as improved 
measurements of charged lepton flavour violation (CLFV), electric dipole moments (EDMs) and 
$(g-2)_{\mu,e}$ will definitely shed light on the question whether NP is present 
below, say, $10\tev$. However, in the coming decades only rare processes will allow us to go beyond $10\tev$. These are
 in particular particle-antiparticle mixings ($\Delta F=2$ processes), rare 
decays of mesons ($\Delta F=1$ processes), CLFV, EDMs  and $(g-2)_{\mu,e}$. As this is an indirect 
search for NP one has to develop special strategies to reach  the Zeptouniverse, that is scales as short as $10^{-21}{\rm m}$ or equivalently energy scales as high as several hundreds of TeV.
The present lecture  discusses some of  such strategies developed in my group at the Technical University in Munich during last ten years. They are summarized in \cite{Buras:2013ooa}  and in my talk at EPS-HEP13 in Stokholm \cite{Buras:2013dda} which I opened with a similar overture as the general situation 
as far as NP is concerned has not changed by much since then. We are still waiting for NP. 

Yet, the rest will include new results and the goals will be  presented 
 in a bit different manner even if some overlap with \cite{Buras:2013dda} is unavoidable.
In any case it is unquestionable that NP beyond the SM exists and it is our 
duty to search for it,  not only through high energy collisions at the LHC, but  in particular through 
rare transitions both in the quark and lepton sectors that we listed above. This is not only because we would like to know which new creatures exist down there, 
but in particular in order to answer many open questions, which I will not repeat here  as they are known to every particle physicist. But let me still 
reemphasize that the issue of {\it the origin of flavour} plays a very important role in 
these efforts for a very obvious reason. In our search for a more fundamental 
theory we need to improve our understanding of flavour. 

While we mentioned above that the SM describes the present data well, there 
are several signals in the present flavour data, which gives us hopes that the increase of precision in experiment and theory in the coming years could indeed provide valuable information about the nature of NP beyond the SM. 
It is strategically useful to list already here these departures from SM 
expectations, write a few lines about the possible NP behind them  and elaborate on some of them later. These are:

{\bf 1.} Anomalies in angular observables in $B_d\to K^*\mu^+\mu^-$ and in the 
branching ratio $\mathcal{B}(B_d\to K^*\mu^+\mu^-)$ that is found to be {\it smaller} than predicted within the SM. A heavy neutral gauge boson $Z^\prime$ or 
leptoquarks with particular transformation properties under the SM gauge group 
are the leading candidates among the ones proposed as the origin of these deviations.

{\bf 2.} Breakdown of lepton flavour universality in  $B^+\to K^+\ell^+\ell^-$ 
with the branching ratio for $B^+\to K^+ e^+e^-$ in agreement with the 
SM but   $\mathcal{B}(B^+\to K^+\mu^+\mu^-)$ found to be {\it smaller} than predicted by the SM \cite{Aaij:2014ora}. Again, a heavy $Z^\prime$ or leptoquarks are the simplest explanations of this behaviour.

{\bf 3.} The ratio
\begin{equation}
\frac{\mathcal{B}(B_d\to \mu^+\mu^-)}{\mathcal{B}(B_s\to \mu^+\mu^-)}=
(4.8\pm 2.2)\left[\frac{\mathcal{B}(B_d\to \mu^+\mu^-)}{\mathcal{B}(B_s\to \mu^+\mu^-)}\right]_\text{SM}
\end{equation}
 is found to be larger than predicted in the SM. Interestingly, as we will see below, while $\mathcal{B}(B_s\to \mu^+\mu^-)$ appears to be {\it smaller} than its SM value, 
$\mathcal{B}(B_d\to \mu^+\mu^-)$ is measured to be larger. This pattern violates the 
predictions not only of the SM, but of the full class of models with minimal 
flavour violation (MFV) where the enhancement (suppression) of one of these 
branching ratios implies uniquely  enhancement (suppression) of the other one.
Here again a $Z^\prime$ can help \cite{Buras:2013ooa}.

{\bf 4.} Moderate tensions between $K^0-\bar K^0$ and $B^0_{s,d}-\bar B^0_{s,d}$ mixing  observables which have been with  us already since 2008 \cite{Lunghi:2008aa,Buras:2008nn}. Typically either the CP asymmetry $S_{\psi K_S}$ 
is above the data or $\varepsilon_K$ is below the data. Moreover, if one increases 
$\vcb$ to get a better result for $\varepsilon_K$, the mass differences $\Delta M_{s,d}$ are shifted above the data. But the uncertainties in the relevant 
$\hat B_i$ parameters are too large to claim the presence of NP. Future lattice  QCD calculations and in particular improved determination of the CKM parameters will clarify this\footnote{As the $\hat B_K$ parameter, relevant for $\varepsilon_K$, is already precisely known and the lattice analyses favour significantly 
lower values of $\vcb$ and $\vub$ than inclusive determinations, from the point of view of lattice QCD, the SM value of $\varepsilon_K$ turns out to be by $3\sigma$ below the data \cite{Bailey:2015tba}.}. This is important as these tensions signal the 
presence of NP beyond MFV. While $Z^\prime$ could also help here, heavy neutral scalars or the reduction of flavour symmetry of MFV models ($U(3)^3$) down to $U(2)^3$ would also work. However, $U(2)^3$ models cannot help with 
the anomalies {\bf 1-3} listed above and in the case of {\bf 1} and {\bf 2} this also applies to heavy neutral scalars.

{\bf 5.} Significant departures from SM expectations in $B_d\to D\tau\nu_\tau$ 
and  $B_d\to D^*\tau\nu_\tau$ signaling possibly the presence of 
new heavy charged scalars or gauge bosons.

{\bf 6.} The $(g-2)_\mu$ anomaly which was with us for more than ten years.

In this lecture I will have nothing to add about the last two tensions to 
what is known already in the literature and I refer to the review \cite{Buras:2013ooa} 
where brief summaries of these two topics together with the relevant literature 
can be found. Similar, I have presently nothing to add to various possible 
departures from SM expectations in non-leptonic decays of $B$ and $D$ mesons, 
partly because theoretical uncertainties in these decays are larger than 
in the processes discussed below. Yet, it should be emphasized that in the flavour precision era, in which the measurements of a multitude of non-leptonic decays will be very much improved, also these decays could be very useful, in particular in the tests of CP violation \cite{Fleischer:2014qla}.

\section{Basic Requirements for Reaching the Zeptouniverse}
The coming ten years (2015-2025) of flavour precision era invites us to  
attempt  an {\it expedition from the Attouniverse to the Zeptouniverse}. For such an 
expedition to have a chance to be successful at least the following 
requirements have to be fullfiled:
\begin{itemize}
\item 
Many precise measurements of many observables.
\item
Precise extraction of CKM parameters from tree level decays which are 
expected to have at most tiny NP contributions\footnote{Recent analyses of  the room left for NP in tree-level decays can be found in \cite{Bobeth:2014rda,Bobeth:2014rra,Brod:2014bfa}.}. Here, the main targets for 
coming years are
\be
\vub, \qquad \vcb,\qquad \gamma
\ee
where $\gamma$, one of the angles in the Unitarity Triangle, is up to the sign 
the complex phase of $V_{ub}$.
\item
Precise lattice QCD calculations of weak decay constants $F_{B_s}$ and $F_{B_d}$,  of various non-perturbative parameters $\hat B_i$ and of form factors 
for various semi-leptonic transitions, in particular for $B\to K(K^*)$ transitions with leptons in the final state and those relevant for the determinations 
of $\vub$ and $\vcb$.  For a review see
 \cite{El-Khadra:2014sha} and references therein.
In the case of non-leptonic decays of mesons lattice QCD 
appears to be less useful and in this case approaches like QCD factorization approach \cite{Beneke:1999br} and those based on flavour symmetries and their breakdown  \cite{Fleischer:2014qla} will continue to play  important roles.
\item
NLO and NNLO QCD corrections and NLO electroweak corrections to various Wilson 
coefficients. Among the tasks listed here I would claim that at least within 
the SM this task has been completed after 26 years of efforts by several theorists (1988-2014). An updated review of these efforts can be found in \cite{Buras:2011we}. I do not think we need more precision here within the SM and these 
calculations are sufficiently demanding that there is no point in doing them 
in extensions of the SM before we know what nature is telling us about NP. 
An exception are tree-level flavour changing neutral currents mediated by 
$Z^\prime$, $Z$ or heavy neutral scalars. Their structure is sufficiently simple so that  NLO QCD corrections to these exchanges could be easily calculated
 \cite{Buras:2012fs,Buras:2012gm}.
\end{itemize}

Concerning the second item one needs the clarification of the discrepancies between inclusive and exclusive determinations of $\vcb$ and $\vub$ from tree-level decays  \cite{Ricciardi:2014aya,Ricciardi:2013cda}. The present status of these determinations is as follows. From 
{\it inclusive} determinations based dominantly on heavy quark expansions one has \cite{Alberti:2014yda,Aoki:2013ldr} 
\be
\vub_\text{incl}=(4.40\pm0.25)\cdot 10^{-3},\qquad \vcb_\text{incl}=(42.21\pm0.78)\cdot 10^{-3}\,,
\ee
whereas from {\it exclusive} determinations based dominantly on formfactors from lattice QCD one has \cite{Bailey:2014tva,Bailey:2014bea,Aoki:2013ldr}
\be
\vub_\text{excl}=(3.72\pm0.14)\cdot 10^{-3},\qquad \vcb_\text{excl}=(39.36\pm0.75)\cdot 10^{-3}\,.
\ee
These differences introduce significant uncertainties in SM predictions for 
rare $K$ and $B_{s,d}$ decays, which have to be decreased by much if we want 
to study NP efficiently. In particular, rare decays $\kpn$ and $\klpn$  and 
$\varepsilon_K$ are 
very sensitive to the value of $\vcb$ but also the branching ratios for 
$B_{s,d}\to \mu^+\mu^-$ and the mass differences $\Delta M_{s,d}$ depend 
quadratically on it. It is likely that this problem will be resolved only by Belle II experiment at SuperKEKB  at the end of this decade, but one should hope 
that further theoretical efforts will tell us what is really going on. From 
the analyses in \cite{Crivellin:2014zpa,Bernlochner:2014ova} it is unlikely that NP, like right-handed charged currents,  are responsible for these discrepancies, but this should be further clarified. 

For the time being one can take a weighted average of these results and scale the errors based on the resulting $\chi^2$, which gives \cite{Buras:2015qea}
\be\label{average}
\vub_{\rm avg} =(3.88\pm0.29)\times 10^{-3}, \qquad  \vcb_{\rm avg}=(40.7\pm1.4)\times 10^{-3}.
\ee
For the CKM angle $\gamma$ the current world average of direct measurements~\cite{Trabelsi:2014} reads
\begin{equation}\label{gamma}
    \gamma = (73.2^{+6.3}_{-7.0})^\circ.
\end{equation}
The fourth element of the CKM matrix is already very well known 
\begin{equation}
    |V_{us}| = 0.2252\pm 0.0009\,.
\end{equation}

Concerning QCD lattice calculations of non-perturbative parameters relevant for 
$\Delta F=2$ transitions and of formfactors entering rare decays like 
$B_d\to K(K^*)\ell^+\ell^-$ and $B_d\to K(K^*)\nu\bar\nu$, where in addition 
to lattice QCD also light-cone sum rules (LCSR) \cite{Straub:2015ica} play an important role, I am 
optimistic that the coming years will bring significant advances. Therefore, I 
am in close contact with lattice and LCSR experts und look up frequently the 
updates  by FLAG \cite{Aoki:2013ldr} and HFAG \cite{Amhis:2012bh}.

In the course of this lecture we will give several examples which 
demonstrate that it is very important to fulfil
the requirements listed above if we want to reach the Zeptouniverse before 
a $100\tev$ collider will be built.

\section{Main Strategy}
\subsection{General View}
Let us then assume that the CKM parameters have been determined with high 
precision and non-perturbative parameters, relevant both for the SM and 
its extentions, have been calculated accurately. Having then precise SM predictions let us assume 
that future precise measurements of various observables have identified a number of deviations from SM predictions so that without any 
doubt we can conclude that some NP is at work. The question then arises 
what kind of NP could be responsible for these deviations. 

Clearly, the most interesting and favourable situation that one could hope
 for, 
would be a direct discovery of new particles at the LHC which would indicate at 
least first steps towards the Zeptouniverse. The interplay of LHC findings with   
quark and lepton flavour data and those on EDMs would be exciting and would teach us a lot, but we cannot exclude at present that the lightest 
 new particles are out of the reach of the LHC.

This would make the life of flavour physicists much harder, but still it 
is our duty to develop efficient tools for the identification of  NP through rare processes, that is through quantum fluctuations.  As summarized 
in \cite{Buras:2013ooa} this will require
\begin{itemize}
\item
many precise measurements of many observables and precise theory,
\item
intensive studies of correlations between many observables in a given extension of the SM with the goal to identify patterns of deviations 
from the SM expectations characteristic for this extension,
\item
intensive studies of  correlations between low energy precision measurements, including electroweak precision tests and the measurements at the highest available energy, that is in the coming decades the measurements of a multitude of observables in proton-proton collisions at the LHC.
\end{itemize}

Now in the search for NP 
 one distinguishes between {\it bottom-up} 
and {\it top-down} approaches. In my view both approaches should be pursued 
but I think that 
in the context of flavour physics and simultaneous 
 exploration of 
short distance physics, both through LHC and high precision experiments, the 
top-down approach is more powerful.  I presented my arguments already in 
\cite{Buras:2013dda} but let me repeat them briefly here.

In the  bottom-up approach
one constructs effective field theories involving 
only light degrees 
of freedom including the top quark and Higgs boson in which the structure of the effective 
Lagrangians is governed by the symmetries of the SM and often other 
hypothetical symmetries. This approach is rather powerful in the case of
electroweak precision 
studies and definitely teaches us something about $\Delta F=2$ 
transitions. In particular, lower bounds on NP scales, depending on the 
Lorentz structure of involved operators, can be derived from the data 
\cite{Bona:2007vi,Isidori:2010kg,Charles:2013aka}.
However, except for the case of  minimal flavour violation (MFV) and closely related 
approaches based on flavour symmetries, the bottom-up approach ceases, 
in my view, to be useful in $\Delta F=1$ decays. Indeed, in this case the appearance of very many operators that are allowed to enter
 the effective Lagrangians with coefficients that are basically 
unknown \cite{Buchmuller:1985jz,Grzadkowski:2010es}, lowers the predictive 
power of theory. In this approach then the correlations between various $\Delta F=2$ and $\Delta F=1$ observables in $K$, $D$, $B_d$ and $B_s$ systems are either not visible or very weak, again except MFV and closely related approaches. Moreover, the correlations between flavour violation in low energy processes, electroweak  precision observables and flavour violation in high energy processes 
are washed out. Again, MFV is among few exceptions. The situation improves when 
only a certain class of processes, like $b\to s \mu^+\mu^-$ and $b\to s\nu\bar\nu$, are considered and the invariance of NP  under the full SM gauge symmetry $SU(3)_c\otimes SU(2)_L\otimes U(1)_Y$ is imposed. This was  
stressed in particular in \cite{Alonso:2014csa} but also in \cite{Hiller:2014yaa,Buras:2014fpa}. Still, as we will discuss below, this approach has limitations in identifying the correct route to short distance scales.

On the other hand
in the top-down 
approach one constructs first 
a specific model with heavy degrees of freedom. For high energy processes,
where the energy scales are of the order of the masses of heavy particles 
one can directly use this ``full theory'' to calculate various processes 
in terms of the fundamental parameters of a given theory. For low energy 
processes one again constructs the low energy theory by integrating out 
heavy particles. The advantage over the bottom-up approach is that now the 
Wilson 
coefficients of the resulting local operators are calculable in terms of 
the fundamental parameters of this theory. In this manner correlations between 
various observables belonging to different mesonic systems and correlations 
between low energy and high-energy observables and also electroweak precision 
tests are possible. Such correlations 
are less sensitive to free parameters than individual observables and 
represent patterns of flavour violation characteristic for a given theory. 
These correlations can in some models differ strikingly from the ones of 
the SM and of the MFV approach.

\begin{figure}[!tb]
\centerline{\includegraphics[width=0.65\textwidth]{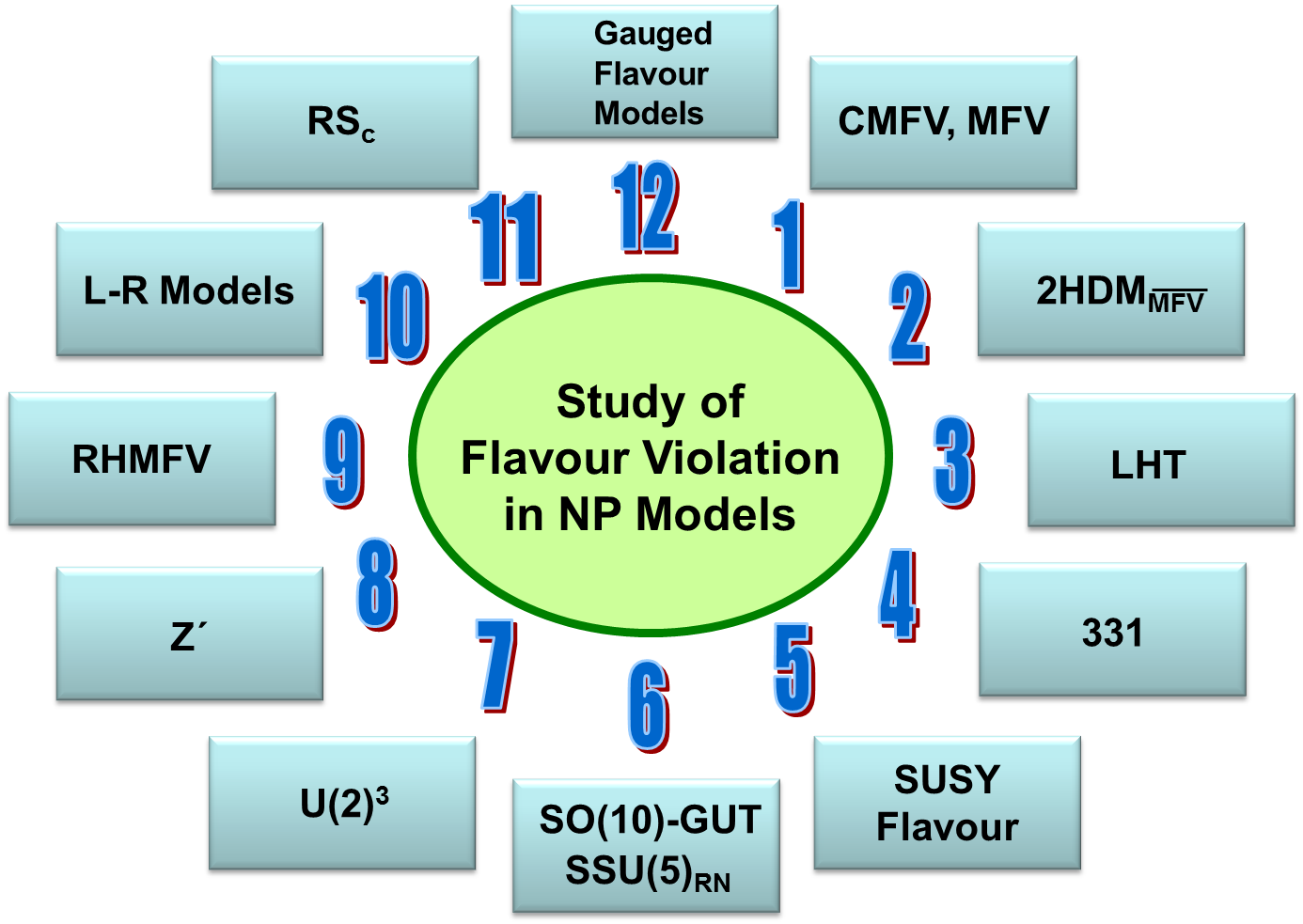}}
\caption{\it Studing multitude of extensions of the Standard Model.}\label{Fig:2}~\\[-2mm]\hrule
\end{figure}

Having the latter strategy in mind I have in the last ten years
investigated together with my 
young collaborators 
flavour violating and CP-violating processes 
in a multitude of models. The names of models analyzed by us until now are collected in Fig.~\ref{Fig:2}. A summary of these studies with brief descriptions of all these models can be found in  \cite{Buras:2010wr,Buras:2012ts,Buras:2013ooa}. Here, I want to summarize some of the lessons gained through these studies and subsequently concentrate on most recent analyses that have been performed by us in  2014 and 2015.

\subsection{Superstars and Stars for 2015-2025 in Quark Flavour Physics}
Yet, before doing this it is useful to list the most promising observables 
in the search for NP. There are many measurements one can do and many 
observables one can calculate, but in my view the following ones will lead the 
discussions and searches for NP in the coming ten years.
\begin{itemize}
\item
$\Delta F=2$ observables:
\be
\Delta M_s, \qquad \Delta M_d, \qquad S_{\psi K_S} \qquad S_{\psi\phi}, \qquad 
\varepsilon_K, 
\ee
with $S_{\psi K_S}$ and $S_{\psi\phi}$ being mixing induced CP-asymmetries that 
are measured in $B_d(\bar B_d)$ decays. In spite of non-leptonic nature of 
the relevant decays, 
these asymmetries have only small hadronic uncertainties 
originating primarly in QCD penguin contributions. On the other hand, 
it is expected that lattice QCD will provide in the next five years 
rather precise values for the hadronic matrix elements relevant for 
$\Delta M_s$, $\Delta M_d$ and $\varepsilon_K$ including the ones present 
in the extensions of the SM. The data on these three quantities are already very precise. The ones for $S_{\psi K_S}$ and $S_{\psi\phi}$ should be precise in the second half of this decade. Their present values are given in (\ref{SDATA}).
\item
Angular observables and branching ratios in the decays
\be
B\to K^*\ell^+\ell^-, \qquad B \to K\ell^+\ell^-\,,
\ee
which are presently stars of flavour physics.
In particular in the case of $B\to K^*\ell^+\ell^-$ the presence 
of a multitude of angular observables could  offer detailed insight 
in the structure of NP. Yet, there are still important issues of long distance 
uncertainties in  $B\to K^*\ell^+\ell^-$ which remain to be clarified. The 
 decays  $B\to K\ell^+\ell^-$ are theoretically cleaner and possible signs 
of the violation of lepton universality seen in the data are very intriguing.
\item
The branching ratios for 
\be
\mathcal{B}(B_{s}\to\mu^+\mu^-), \qquad \mathcal{B}(B_{d}\to\mu^+\mu^-)\,.
\ee
The weak decay constants entering these branching ratios should be 
known in the coming years with the accuracy of $(1-2)\%$. The uncertainty 
in $\vts$ and $\vtd$ can be totally removed within CMFV by relating these branching 
ratios to $\Delta M_s$ and $\Delta M_d$, respectively \cite{Buras:2003td}. In other scenarios 
the corresponding relation does not fully remove this uncertainty but 
can reduce it significantly. The $\hat B_{s,d}$ parameters entering 
these relations should be known within $2\%$ in this decade from 
lattice QCD.
\item
The branching ratios
\be
\mathcal{B}(B\to K \nu \bar \nu), \qquad \mathcal{B}(B\to K^* \nu \bar \nu), \qquad \mathcal{B}(B\to X_s \nu \bar \nu).
\ee
Even if these branching ratios are sensitive to form factor uncertainties, 
which should be decreased significantly in this decade, they are less 
subject to long distance effects present in $b\to s\mu^+\mu^-$. Moreover, 
they offer powerful means to study the effects of right-handed currents \cite{Colangelo:1996ay,Buchalla:2000sk,Altmannshofer:2009ma,Buras:2012jb,Biancofiore:2014uba,Buras:2014fpa}.
\item
The branching ratios for 
\be
\mathcal{B}(\kpn), \qquad \mathcal{B}(\klpn)\,.
\ee
These two branching ratios are basically free from hadronic uncertainties 
and the ones in the charm contribution to $\kpn$ can be reduced in the 
future through lattice QCD. The very strong dependence on $\vcb$ of both branching ratios and on $\vub$ in the case of $\klpn$  will 
remain an issue for some time. A recent analysis in \cite{Buras:2015qea} demonstrates it explicitly. On the other hand the triple correlation 
between these two branching ratios and  $S_{\psi K_S}$ within the SM and CMFV 
models is practically 
free from this dependence \cite{Buchalla:1994tr,Buras:2001af}. In this 
context one should mention the ratio $\epe$ that is very sensitive to NP 
effects. Unfortunately, the status of hadronic matrix elements relevant for 
the calculation of $\epe$ is far from being satisfactory. While the ones relevant for electroweak penguin contributions are known within the accuracy of $5\%$ 
from lattice QCD \cite{Blum:2015ywa}, the ones of QCD penguins are known only in the large $N$ limit of QCD \cite{Buras:2014maa}. From present perspective it is unlikely that $\epe$ will become 
the star of flavour physics in the coming years, but could become it around 
2020 and in the next decade.
\item
\be
\mathcal{B}(B^+\to\tau^+\nu_\tau), \qquad \mathcal{B}(B\to D \tau\nu_\tau), \qquad \mathcal{B}(B\to D^*\tau\nu_\tau)\,,
\ee
which play important role in searching for NP effects mediated by charged scalars and gauge bosons \cite{Nierste:2008qe,Kamenik:2008tj}. Moreover, the present data from BABAR  \cite{Lees:2012xj} on the last two 
show the largest departures from the SM in flavour physics. We refer to 
\cite{Fajfer:2012vx,Fajfer:2012jt,Crivellin:2012ye,Crivellin:2013wna,Ko:2012sv,Crivellin:2013mba,Fajfer:2013aw} for details.
\end{itemize}
\subsection{Superstars and Stars for 2015-2025 in Lepton Flavour Physics} 
There is no question about that also lepton flavour physics will play a very 
important role in identifying NP beyond the SM. In particular, the following decays should provide a deep insight into the dynamics at short distance scales:

\begin{itemize}
\item
First of all the decays 
\be
\mu\to e \gamma, \qquad \tau\to e\gamma, \qquad \tau\to \mu\gamma,
\ee
that are governed by dipole operators. The improved bound on the first one from 
MEG  \cite{Adam:2013mnn}
\be
 \mathcal{B}(\mu\to e\gamma)\le (5.7)\times 10^{-13}\,
\ee
puts significant constraints on the parameters of various extensions of the 
SM. The improvement on the other two decays is expected from Belle II and LHCb.
\item
Next come
\be
\mu^-\to e^-e^+e^-, \qquad \tau^-\to\mu^-\mu^+\mu^-, \qquad \tau^-\to e^-e^+e^-\,.
\ee
These decays are very interesting as they are strongly correlated with 
$\mu\to e \gamma$, $\tau\to e\gamma$ and $\tau\to \mu\gamma$ and these correlations are different for different models. 
\item
Also the four decays 
\be
\tau^-\to e^-\mu^+e^-, \qquad \tau^-\to\mu^- e^+\mu^-,
\ee
and
\be
\tau^-\to\mu^-e^+e^-, \qquad \tau^-\to e^-\mu^+\mu^-
\ee
will enrich the search for NP.
\item
$\mu-e$ conversion in nuclei, even if subject to hadronic uncertainties, could 
become the star of lepton flavour physics at the end of this decade.  The dedicated J-PARC experiment PRISM/PRIME  should reach a sensitivity of $\ord(10^{-18})$ \cite{Barlow:2011zza}. Also, semi-leptonic $\tau$ decays like $\tau\to\pi\mu e$ should not be forgotten.
\end{itemize}

For further detailed review of LFV see \cite{Raidal:2008jk,Feldmann:2011zh,Ibarra:2010zz}. An experimenter's guide for charged LFV  can be found in 
 \cite{Bernstein:2013hba}.

\subsection{Interplay of Quark and Lepton Flavour Violation} 
Of special interest are decays that proceed through both quark flavour and 
lepton flavour violating transitions. These are in particular
\be
K_{L,S}\to\mu e, \qquad K_{L,S}\to\pi^0\mu e \,,
\ee
\be 
B_{d,s} \rightarrow \mu e, \qquad B_{d,s} \rightarrow \tau e, \qquad B_{d,s} \rightarrow \tau \mu\,
\ee
and 
\be
B_d\to K^{(*)}\tau^\pm\mu^\mp, \qquad B_d\to K^{(*)}\mu^\pm e^\mp\,.
\ee
A natural mechanism responsible for such transitions are tree-level exchanges 
of leptoquarks \cite{Varzielas:2015iva} or tree-level $Z^\prime$ exchanges 
 \cite{Crivellin:2015era}. 
But they can also be generated at one-loop level, an example 
being LHT model  \cite{Blanke:2007db}.
\subsection{Electric Dipole Moments and $(g-2)_{e,\mu}$}
Even if these observables are flavour conserving they put strong bounds on 
extensions of the SM. The $(g-2)_{\mu}$ anomaly found at Brookhaven should 
be clarified by Fermilab at the end of this decade. A recent review about EDMs can be found in \cite{Engel:2013lsa} which updates the review in \cite{Pospelov:2005pr}. See also \cite{Batell:2012ge}. 

After this collection of most important decays let us enter some details.

\section{The Power of Correlations between Flavour Observables}
\subsection{Preliminaries}
In studying correlations between various decays it is important to remember that
\cite{Blanke:2014asa}
\begin{itemize}
\item
Correlations between decays of different mesons test the flavour structure 
of couplings or generally flavour symmeteries.
\item
Correlations between decays of a given meson test the Dirac structure of 
couplings.
\end{itemize}
We will first look at the first correlations by comparing those within 
MFV models based on $U(3)^3$ flavour symmetry with the ones present in 
models with $U(2)^3$ flavour symmetry. To this end we will assume that 
in the latter case, similar to MFV, only the left-handed couplings are 
relevant.
\subsection{CMFV and MFV}
These models are based on flavour $U(3)^3$ symmetry and their most striking 
predictions are:
\begin{itemize}
\item
No new sources of flavour and CP violation (excluding flavour blind phases) imply \footnote{Our definition 
of $S_{\psi\phi}$ differs by sign from the one used by LHCb and HFAG.}.
\be\label{CPSM}
S_{\psi K_S}=\sin 2\beta, \qquad S_{\psi\phi}= S_{\psi\phi}^\text{SM}=0.036\pm0.002
\ee
\item
Stringent correlations between $K$, $B_d$ and $B_s$ systems and in 
particular between $\Delta F=2$ and $\Delta F=1$ observables. 
\item
For fixed CKM parameters determined in tree-level decays, $|\varepsilon_K|$, 
 $\Delta M_s$ and $\Delta M_d$, if modified,  can only be {\it enhanced} 
 relative to SM predictions  \cite{Blanke:2006yh}. Moreover, this 
 happens in a correlated manner \cite{Buras:2012ts,Buras:2000xq}.
 The implications of this property are rather powerful. Finding in the future SM 
prediction for one of these three observables  above 
its experimental value, will signal the presence of non-CMFV interactions.
\item
Absence of right-handed charged currents.
\end{itemize}

\begin{figure}[!tb]
 \centering
\includegraphics[width = 0.6\textwidth]{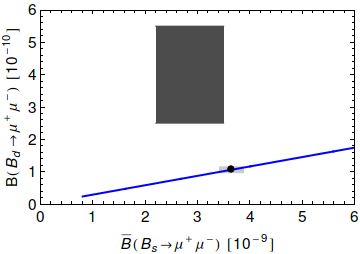}
\caption{\it $\mathcal{B}(B_d\to\mu^+\mu^-)$ vs $\overline{\mathcal{B}}(B_s\to\mu^+\mu^-)$ 
  in models 
with CMFV. SM is represented by the light grey area with black dot. Dark gray
 region: Overlap of  exp 1$\sigma$
 ranges for
 $\overline{\mathcal{B}}(B_s\to\mu^+\mu^-) = (2.8\pm0.7)\cdot 10^{-9}$ and 
$\mathcal{B}(B_{d}\to\mu^+\mu^-) =(3.9^{+1.6}_{-1.4})\times 10^{-10}$. Update of 
\cite{Buras:2013ooa}.}\label{fig:BdvsBs}~\\[-2mm]\hrule
\end{figure}

Let me recall some of these correlations as they could be soon relevant. 
The first two are the ones in models with constrained Minimal Flavout Violation 
(CMFV)  \cite{Buras:2000dm,Buras:2003jf}
\begin{equation}\label{CMFV5}
 \frac{\mathcal{B}(B_s\to\mu^+\mu^-)}{\mathcal{B}(B_d\to\mu^+\mu^-)}=
 \frac{\tau({B_s})}{\tau({B_d})}\frac{m_{B_s}}{m_{B_d}}
 \frac{F^2_{B_s}}{F^2_{B_d}}
 \left|\frac{V_{ts}}{V_{td}}\right|^2,
 \end{equation}
and \cite{Buras:2003td}\footnote{As emphasized in \cite{Hurth:2008jc} the 
dependence of the ratio of branching ratios in (\ref{CMFV5}) on the elements 
of the CKM is more general than CMFV and applies to MFV at large \cite{Chivukula:1987py,Hall:1990ac,D'Ambrosio:2002ex}.}
\be\label{CMFV6}
 \frac{\mathcal{B}(B_{s}\to\mu^+\mu^-)}{\mathcal{B}(B_{d}\to\mu^+\mu^-)}
 =r\frac{\hat B_{d}}{\hat B_{s}}
 \frac{\tau( B_{s})}{\tau( B_{d})} 
 \frac{\Delta M_{s}}{\Delta M_{d}}=r\,(34.5\pm0.8),  \qquad 
\frac{\hat B_{d}}{\hat B_{s}}=0.99\pm0.02
\ee
where the departure of $r$ from unity measures the effects which go beyond CMFV. 
This {\it golden} relation  between $\Delta M_{s,d}$ and $B_{s,d}\to\mu^+\mu^-$ 
does not 
 involve $F_{B_q}$ and CKM parameters. Consequently it contains 
 smaller hadronic and parametric uncertainties than (\ref{CMFV5}). It involves
 only measurable quantities except for the ratio $\hat B_{s}/\hat B_{d}$
  that is known from  lattice calculations with impressive 
accuracy of roughly $\pm 2\%$ \cite{Carrasco:2013zta} as given in (\ref{CMFV6}).
 Consequently the r.h.s of this equation is already  rather precisely 
 known and this precision will be improved within this decade. 
 This would allow us to identify possible NP in 
 $B_{s,d}\to \mu^+\mu^-$ decays and also in $\Delta M_{s,d}$  even if it was only at the level of $20\%$ of 
 the SM contributions. This is rather unique in the quark flavour physics and 
only the decays $\kpn$ and $\klpn$ can compete with this precision. 

In fact, in the case of the latter decays one can find a theoretically clean 
relation between the branching ratios for $\kpn$ and $\klpn$ and $\sin 2\beta$
that practically does not depend on CKM uncertainties, in particular $\vcb$, 
$\vub$ and $\gamma$, even if  the individual branching ratios for these decays 
are rather sensitive to the values of these parameters as stressed recently 
in \cite{Buras:2015qea}. We refer to \cite{Buchalla:1994tr,Buras:2001af}, Section~\ref{sec:8} in this lecture and \cite{Buras:2015qea} for details.

The most recent data  on $B_s\to\mu^+\mu^-$
 from LHCb and CMS collaborations give first indications that NP 
 contributions to $B_s\to\mu^+\mu^-$  are much smaller than the 
SM contribution itself. On the other hand, the data on $B_d\to\mu^+\mu^-$ exhibit some departure from SM expectations, but we have to wait for improved data in order to see whether NP is at work here. 
We compare the relation (\ref{CMFV6}) with present data in  Fig.~\ref{fig:BdvsBs}, where we included $\Delta\Gamma_s$ effects in $B_s\to\mu^+\mu^-$. 

 The most recent prediction in the SM  that includes NNLO QCD corrections  \cite{Hermann:2013kca}  and 
NLO electroweak corrections  \cite{Bobeth:2013tba}, put together in \cite{Bobeth:2013uxa}, and the 
most recent averages from the combined analysis of CMS and LHCb \cite{CMS:2014xfa} are given as follows: 
\be\label{LHCb2}
\overline{\mathcal{B}}(B_{s}\to\mu^+\mu^-)_{\rm SM}= (3.65\pm0.23)\cdot 10^{-9},\quad
\overline{\mathcal{B}}(B_{s}\to\mu^+\mu^-) = (2.8^{+0.7}_{-0.6}) \times 10^{-9}, 
\ee
\be\label{LHCb3}
\mathcal{B}(B_{d}\to\mu^+\mu^-)_{\rm SM}=(1.06\pm0.09)\times 10^{-10}, \quad
\mathcal{B}(B_{d}\to\mu^+\mu^-) =(3.9^{+1.6}_{-1.4})\times 10^{-10}. \quad
\ee
The ``bar'' in the case of $B_{s}\to\mu^+\mu^-$ indicates that $\Delta\Gamma_s$ 
effects \cite{DescotesGenon:2011pb,DeBruyn:2012wj,DeBruyn:2012wk} have been taken into account.

Clearly in the case of $B_d\to\mu^+\mu^-$ 
large deviations from SM prediction are still possible. But in the case of
$\overline{\mathcal{B}}(B_{s}\to\mu^+\mu^-)$
deviations by more than $30\%$ from its SM value seem rather unlikely. Yet, 
the reduction of the error in the SM prediction down to $3-4\%$ is still possible and this would allow to see NP at the level of $20\%$ provided the 
measurements improve.

We observe that the data for $\overline{\mathcal{B}}(B_{s}\to\mu^+\mu^-)$ 
are by $1.2\sigma$  lower than the SM prediction. Yet, at this stage I would 
like to express one warning. The authors in \cite{Bobeth:2013uxa} used the inclusive value for 
$\vcb$ in obtaining quoted result. If they had used the exclusive one, the 
central value for the branching ratio would move down to $3.1\times 10^{-9}$, 
that fully overlaps with the data.

In CMFV \cite{Buras:2000dm} and MFV at large \cite{D'Ambrosio:2002ex}, that are both 
based on the $U(3)^3$ flavour symmetry, the measurement of   
the mixing induced asymmetry  $S_{\psi K_S}$ together with $\Delta F=2$ constraints and the unitarity 
of the CKM implies that the analogous asymmetry in the $B_s^0-\bar B_s^0$ 
system, $S_{\psi\phi}$, is very small. See (\ref{CPSM}). Presently, 
the data give 
\be\label{SDATA}
S_{\psi K_S}= 0.679\pm 0.020,\qquad S_{\psi\phi}= 0.010\pm 0.039, \,
\ee
where the first number comes from PDG and the second from the most recent 
analysis of the LHCb \cite{Aaij:2014zsa} which dominates this determination.
Although $S_{\psi\phi}$ is found to be small  it could still significantly differ from its SM value, in particular if it had negative sign. We are looking forward to new world averages of these asymmetries in the coming years.

Now, the SM faces the following problem. In order to reproduce the data 
on $S_{\psi K_s}$ the value of $\vub$ has to be close to its  exclusive 
determination, but then $\varepsilon_K$ turns out to be too small to agree 
with very precise data \cite{Buras:2008nn}. The solution to this problem is a large value
of $\vcb$ in the ballpark of its inclusive determination but then, 
with present lattice input, the SM values of  $\Delta M_s$ and $\Delta M_d$ are 
above the data. 
 Going beyond the SM, but staying withing CMFV, allows to 
improve the agreement of $\varepsilon_K$ with the data by increasing the 
box function $S$ above its SM value. This is natural within CMFV models as 
stated above: $S$ and $\varepsilon_K$ can only be increased  \cite{Blanke:2006yh}. But this function 
also enters  $\Delta M_s$ and $\Delta M_d$  and they are again increased above 
their experimental values. As analyzed in detail in \cite{Buras:2013raa} only 
for specific values of non-perturbative parameters entering $\Delta M_{s,d}$ 
can SM and CMFV be saved. 

In fact this pattern is fully consistent with the most recent UTfitters result 
in \cite{Bevan:2014cya} where the determined values of their coefficients 
$C_{B_d}=0.81\pm0.12$ and $C_{B_s}=0.87\pm0.09$, while consistent with unity 
as obtained within the SM, indicate that the data favour NP that suppresses  $\Delta M_{s,d}$ which is impossible within CMFV  \cite{Blanke:2006yh}.
In summary, it appears as the correlation between 
$\varepsilon_K$, $\Delta M_{s,d}$ and $S_{\psi K_S}$ could turn out to be  a problem for SM and CMFV, but clear cut conclusions can only be reached when the precision on lattice QCD calculations improves.

On the other hand, if one day the value of $\vub$ above $0.0040$ will be the 
correct one, then the predicted value of $S_{\psi K_S}$ will be above the 
data and without  new CP-violating phases we will not be able to 
reproduce its value in (\ref{SDATA}) \cite{Lunghi:2008aa}. 

My personal expectations are that the exclusive determinations of $\vub$ and 
$\vcb$ will win and that NP in $\varepsilon_K$ but not $S_{\psi K_S}$ will 
be required. While in this case CMFV or MFV will still remain to be a valid 
framework, their future will depend on the precise values of hadronic matrix 
elements entering $\Delta M_s$ and $\Delta M_d$ and on other observables listed 
previously.

\subsection{$U(2)^3$ Symmetry}

\begin{figure}[!tb]
 \centering
\includegraphics[width = 0.6\textwidth]{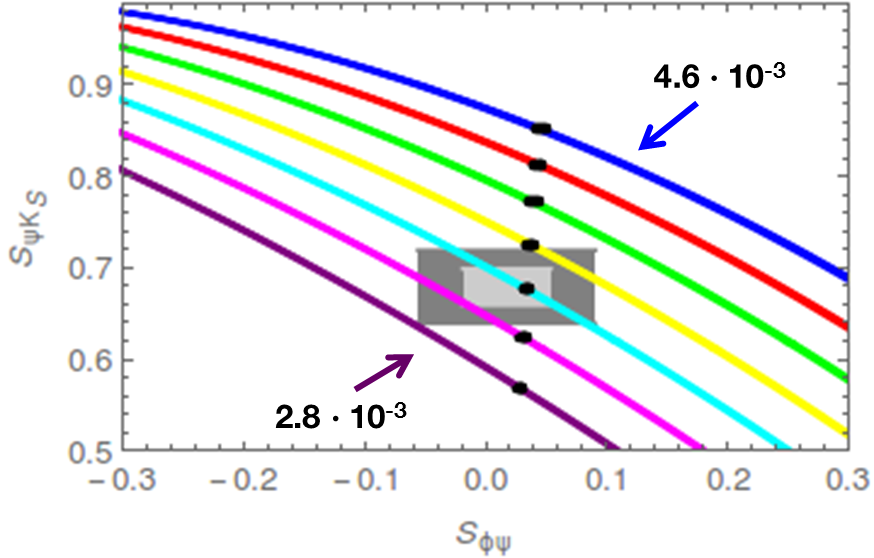}
\caption{ \it $S_{\psi K_S}$ vs. $S_{\psi \phi}$ in  models with 
$U(2)^3$ symmetry for different values of $\vub$ and $\gamma\in[58^\circ,78^\circ]$. From top to bottom: $\vub =$ $0.0046$ (blue), $0.0043$
(red), $0.0040$ (green),
$0.0037$ (yellow), $0.0034$ (cyan), $0.0031$ (magenta), $0.0028$ (purple). Light/dark gray: experimental $1\sigma/2\sigma$ region. Update of \cite{Buras:2012sd}. }\label{fig:SvsS}~\\[-2mm]\hrule
\end{figure}

The tensions between $\varepsilon_K$ and $\Delta M_{s,d}$ on one hand and 
between  $\varepsilon_K$ and $S_{\psi K_S}$ on the other hand originate 
in the strict correlations between $K$, $B_d$ and $B_s$ system present 
in the SM and CMFV models. One elegant solution to this problem is the 
reduction of the flavour symmetry down to $U(2)^3$ \cite{Barbieri:2011fc,Barbieri:2012uh,Crivellin:2011fb}. In this case, the flavour symmetry is only between 
two light quark generations with the following important modifications 
relative to the $U(3)^3$ case:
\begin{itemize}
\item
The formulae in (\ref{CPSM}) are modified to
\be\label{CPSMU2}
S_{\psi K_S}=\sin (2\beta+\varphi^\text{new}), \qquad 
S_{\psi\phi}= \sin (2|\beta_s|-\varphi^\text{new}),
\ee
where $\beta_s=-1^\circ$ is  up to the sign the phase of $-V_{ts}$ and $\varphi^\text{new}$ is a new phase.
\item
While the correlations between $\Delta M_s$ and $\Delta M_d$ and the 
relation in (\ref{CMFV6}) remains true, the correlation between $\varepsilon_K$ 
and $\Delta M_{s,d}$ is broken.
\end{itemize}

Therefore, as seen in the plots in \cite{Barbieri:2014tja}, the tensions discussed above are avoided in this NP scenario but
 as  pointed out in \cite{Buras:2012sd} in the simplest versions of these models in which this symmetry 
is broken minimally, there is a stringent triple correlation
 $S_{\psi K_S}-S_{\psi\phi}-|V_{ub}|$  that  constitutes 
an important test of these models. 
 We 
show this correlation in Fig.~\ref{fig:SvsS}  for  $\gamma$ between $58^\circ$ and $78^\circ$. The latter dependence is very weak and is represented by the thickness of the lines.  Note that in a $U(2)^3$ symmetric world, $\vub$ could be 
determined with very small hadronic uncertainties by simply measuring $S_{\psi\phi}$ and $S_{\psi K_S}$. However, it is more interesting to extract $\vub$ from tree level decays and check whether this triple correlation is respected by nature. 

While at the time of the analysis in \cite{Buras:2012sd} still significant 
departures of $S_{\psi\phi}$ from its SM value were allowed, the improved data 
on this asymmetry indicate that  $\vub\approx (3.4\pm0.3)\times 10^{-3}$ 
is favoured. This value is in perfect agreement with the most recent exclusive determinations of $\vub$.
   Whether models with $U(2)^3$ symmetry will get some problems here will 
depend on the future determinations of the three quantities 
in question.

But there is still another  important point. In 
this simple scenario the relation (\ref{CMFV6}) is still valid \cite{Buras:2012sd} even if the branching ratios and $\Delta M_{s,d}$ can all differ  from their SM values. This means that if the experimental grey area in Fig.~\ref{fig:BdvsBs} will not move in the future,  but  will decrease in size,  
 the breakdown of $U(2)^3$ symmetry has to be either more involved or we have to look for other alternatives for NP. This brings us to a more general 
study of correlations between flavour observables.

\section{Correlations between Flavour Observables in Models with Tree Level FCNCs}
\subsection{Generalities}
During the last two years we have performed  general analyses of 
 flavour observables in models in which FCNC processes are mediated at tree-level by neutral gauge bosons \cite{Buras:2012jb,Buras:2013td,Buras:2013qja} and neutral scalars or pseudoscalars \cite{Buras:2013uqa,Buras:2013rqa}. In addition we 
have made detailed analyses of FCNCs within the 3-3-1 models \cite{Buras:2012dp,Buras:2013dea,Buras:2014yna}, in particular in 
view of the $B_d\to K^*\mu^+\mu^-$ anomalies and the new experimental results on $B_s\to\mu^+\mu^-$ \cite{CMS:2014xfa}. A review of these analyses can be found in \cite{Buras:2013ooa,Buras:2013dda}, where also the references to other papers written in 2013 and 2014 can be found. Our main goal in these papers was to identify 
correlations between several  flavour observables that have not been presented in the past. Here, I will mainly concentrate on our analyses 
done in 2014. But first let me recall why the correlations in $Z^\prime$ models 
are very instructive. The point is that the structure of such NP contributions is simple. Indeed, 
a  tree level contribution to a $\Delta F=2$ transition, like particle-antiparticle mixing, mediated by a gauge boson $Z^\prime$ is described by the amplitude
\be\label{FCNC1}
\mathcal{A}(\Delta F=2)=a \bar\Delta_B^{ij}(Z^\prime)\bar\Delta_C^{ij}(Z^\prime), 
\qquad \bar\Delta_B^{ij}(Z^\prime)=\frac{\Delta_B^{ij}(Z^\prime)}{M_{Z^\prime}},
\ee
where $\Delta_{B,C}^{ij}$  with $(B,C)=(L,R)$ are left-handed or right-handed 
couplings of $Z^\prime$ to quarks with $(i,j)$ equal 
to $(s,d)$, $(b,d)$ and $(b,s)$ for $K^0$, $B^0_d$ and $B^0_s$ meson system, 
respectively. The overall flavour independent factor $a$ is a numerical constant that generally depends 
on $L$ and $R$ but we suppress this dependence.
If we assume that only left-handed or right-handed couplings 
are present or that left-handed and right-handed couplings are either equal 
to each other or differ by sign, then this amplitude for a fixed 
$(i,j)$ is described by only two parameters, the magnitude and the phase of 
the reduced coupling $\bar\Delta_{B}^{ij}$.

On the other hand, a tree-level amplitude for a $\Delta F=1$ transition like 
a leptonic or semi-leptonic decay of a meson with $\mu\bar\mu$  or $\nu\bar\nu$  appearing in the final state has the structure
\be\label{FCNC2}
\mathcal{A}(\Delta F=1)=b\bar\Delta_B^{ij}(Z^\prime)\bar\Delta_D^{\mu\bar\mu}(Z^\prime), \qquad 
\bar\Delta_D^{\mu\bar\mu}(Z^\prime)=
\frac{\Delta_D^{\mu\bar\mu}(Z^\prime)}{M_{Z^\prime}},
\ee
with $\bar\Delta_B^{ij}(Z^\prime)$ being the same quark couplings as in (\ref{FCNC1}) and $b$ is again an overall factor. $D=(A,V)$ distinguishes between 
 axial-vector 
and vector coupling to muons. For $\nu\bar\nu$ the couplings are chosen to be left-handed: $\Delta_L^{\nu\bar\nu}(Z^\prime)$.
Clearly the same formulae with different values of couplings and the factors 
$a$ and $b$ apply to a tree-level 
exchange of $Z$, a heavy pseudoscalar $A$ and a heavy scalar $H$.

Now we can constrain the $\Delta_B^{bs}(Z^\prime)$ couplings by the data on 
$\Delta M_s$ and 
the CP-asymmetry $S_{\psi\phi}$ and the couplings $\Delta_B^{bd}(Z^\prime)$ by the  data on 
$\Delta M_d$ and the CP-asymmetry $S_{\psi K_S}$. In the case of $\Delta_B^{sd}(Z^\prime)$ we have mainly $\varepsilon_K$ at our disposal as $\Delta M_K$ 
 having significant hadronic uncertainties provides much weaker 
constraint than $\varepsilon_K$   in the models in question.

Once these constraints on the magnitude and the phase of 
new couplings are imposed and the allowed values 
are used for the predictions for rare decays it is evident that correlations 
between various observables are present, although the correlations between the 
decays with $\nu\bar\nu$ and those with $\mu^+\mu^-$ in the final state require  more information, except FCNCs mediated by $Z$ boson, where these couplings 
are known. More about it later.

 It is particularly interesting that
the pattern of these correlations depends on whether a gauge boson,  a scalar 
or pseudoscalar mediates the FCNC transition. As the scalar contributions cannot interfere with SM contributions, only enhancements of branching ratios are possible in this case. A tree-level gauge boson contribution and pseudoscalar 
contribution interfer generally with the SM contribution, but the resulting 
correlations between observables have different pattern because of the $i$ 
in the coupling 
$i\gamma_5$ of a pseudoscalar to leptons. Detailed analytic explanations of these differences and the corresponding plots can be found in \cite{Buras:2013rqa}.

\subsection{Left-handed and Right-handed Couplings in Tree-Level FCNCs}
$\Delta F=2$ transitions provide a useful information about short distance 
dynamics and  played already very important role in constraining the 
extensions of the SM in the last thirty years. Yet, I would like to stress 
one limitation of such tests of NP. Consider a contribution from a tree-level 
 heavy neutral gauge boson to $\varepsilon_K$. Assuming that both left-handed 
and right-handed couplings to quarks are involved, the general structure of 
this contribution is as follows:
\be\label{LRstructure}
\Delta\varepsilon_K= \text{Im}(a g^2_L+ a g_R^2 + b g_Lg_R)
\ee
where $a$ and $b$ are real and the couplings $g_L$ and $g_R$ can be complex. Note the equality of the coefficients in the first two terms which follows form 
the vectorial structure of QCD interactions.
Indeed the coefficients $a$ and $b$ encode the information about hadronic matrix elements and QCD effects, in particular renormalization group effects. Typically 
$b\approx 150 a$. Analogous formulae can be written for $\Delta M_{s,d}$ but 
in this case $b\approx 7 a$.

As the expression in (\ref{LRstructure}) is symmetric under the interchange of 
left and right, even if $\Delta\varepsilon_K$ will indeed be found experimentally to be non-zero, it will not be possible to decide on the basis of $\varepsilon_K$ alone whether left-handed couplings or right-handed couplings or both couplings are responsible for this signal of NP. The same comment applies to $\Delta M_{s,d}$. But the case of 
 $\Delta F=1$ transitions, in particular of rare $K$ and 
rare $B_{s,d}$ decays, is different and they are crucial for getting deeper 
insight into the structure of the dynamics at very short distance scales.

Indded, let us consider a number of prominent rare decays and divide them 
into two classes:

{\bf Class A:} Decays that are governed by vector ($V=\gamma_\mu$) quark couplings. These are for instance
\be
\kpn, \qquad \klpn, \qquad B\to K\nu\bar\nu, \qquad B\to K \mu^+\mu^-.
\ee
In this case the change from left-handed to right-handed quark couplings does not 
introduce any change of the sign of NP contribution relatively to the SM one.

{\bf Class B:} Decays that are governed by axial-vector ($A=\gamma_\mu\gamma_5$) quark couplings. These are for instance
\be
K_L\to\mu^+\mu^-, \qquad B\to K^*\nu\bar\nu, \qquad B_{s,d}\to \mu^+\mu^-, \qquad B_d\to K^* \mu^+\mu^-.
\ee
In this case the change from left-handed to right-handed couplings implies
 the sign flip of NP contribution relatively to the SM one. Strictly speaking 
in the case of  $B\to K^*\nu\bar\nu$ and $B_d\to K^* \mu^+\mu^-$ this rule only  applies if the contributions from the longitudinal and parallel transversity components dominate. For perpendicular component there is no sign flip.

Thus if there is a correlation between two observables belonging to class A 
and B in the presence of left-handed couplings, it is changed into anti-correlation when right-handed couplings are at work. This difference allows then 
to probe whether one deals with left-handed or right-handed couplings. Of 
course if both left-handed and right-handed couplings are involved the 
structure of correlations is modified, but still studying it one can in 
principle extract the relative size of these couplings from the data. Moreover, 
if there is a correlation or anti-correlation of two observables belonging to 
one class, the flip of sign of $\gamma_5$ will not have an impact on these 
relations, but can of course have an impact on whether a given observable is 
suppressed or enhanced relative to the SM prediction.
A graphical representation of these properties are the DNA charts \cite{Buras:2013ooa} which we will briefly discuss now.

\section{Towards the New SM with the Help of DNA-Charts}
\begin{figure}[!tb]
\centerline{\includegraphics[width=0.65\textwidth]{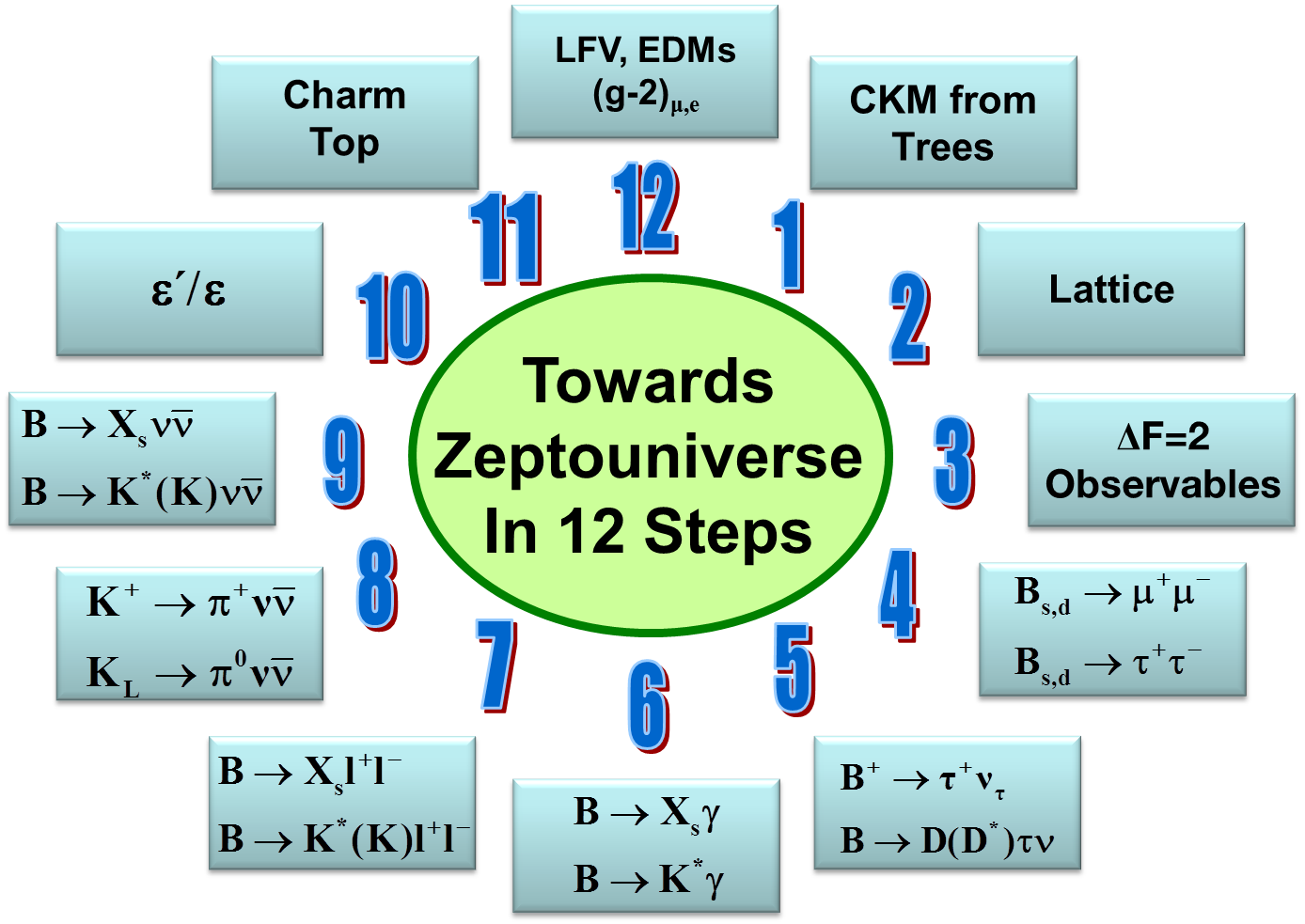}}
\caption{\it Towards the Zeptouniverse in 12 Steps.}\label{Fig:1}~\\[-2mm]\hrule
\end{figure}

The identification of NP indirectly will require many measurements. Fortunately,  the coming years should provide important experimental input  
for this goal. In Fig.~\ref{Fig:1} (slightly different from the one in \cite{Buras:2013ooa}) we collect those processes that we think will play the dominant role 
in testing the short distance structure during the second run of the LHC 
and in the search for NP beyond the LHC reach.  In addition to 
 $\Delta F=2$ processes and rare $K$ and $B_{s,d}$ decays we added other 
measurements not related to quark physics that we already mentioned previously. In fact, simultaneous study of the 
outcome of direct searches for NP at the LHC, electroweak precision tests of 
BSM models suggested for the explanations of anomalies in the quark sector, 
charged lepton flavour violation, EDMs and $(g-2)_{\mu,e}$ could indeed 
help us in the identification of the correct route to the Zeptouniverse and 
allow us to construct a more fundamental theory than the SM.

But as far as quark physics is concerned very important are future precise determinations of CKM parameters and those of 
non-perturbative parameters from lattice QCD. This would allow us to reduce or 
even remove the main uncertainties that are present in the SM predictions for various FCNC observables.
In fact as shown in \cite{Buras:2013qja,Buras:2013raa} in explicit terms
the pattern of NP required for the explanations of deviations from SM predictions depends crucially on this input.

\begin{figure}[!tb]
\centering
\includegraphics[width = 0.49\textwidth]{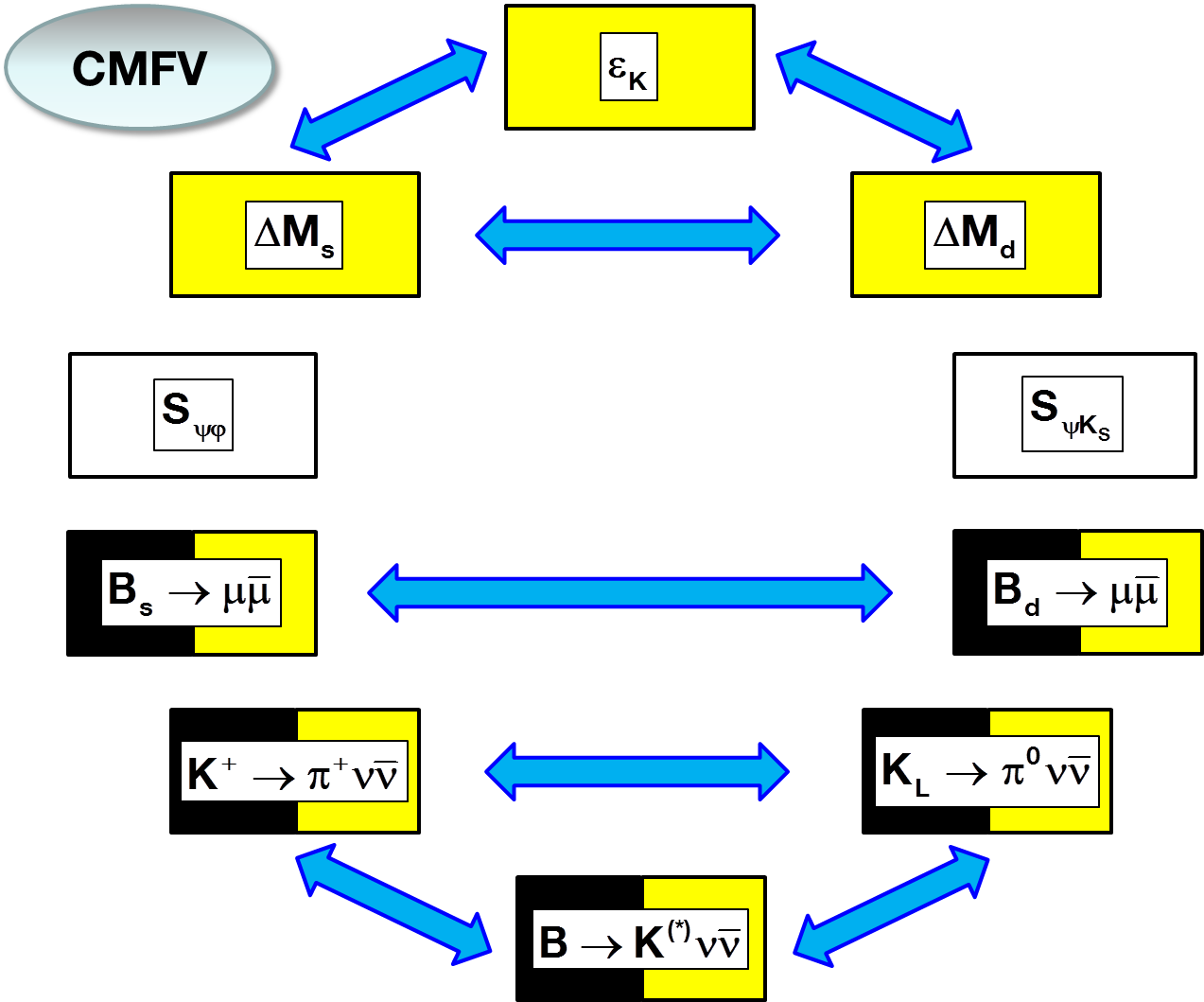}
\includegraphics[width = 0.49\textwidth]{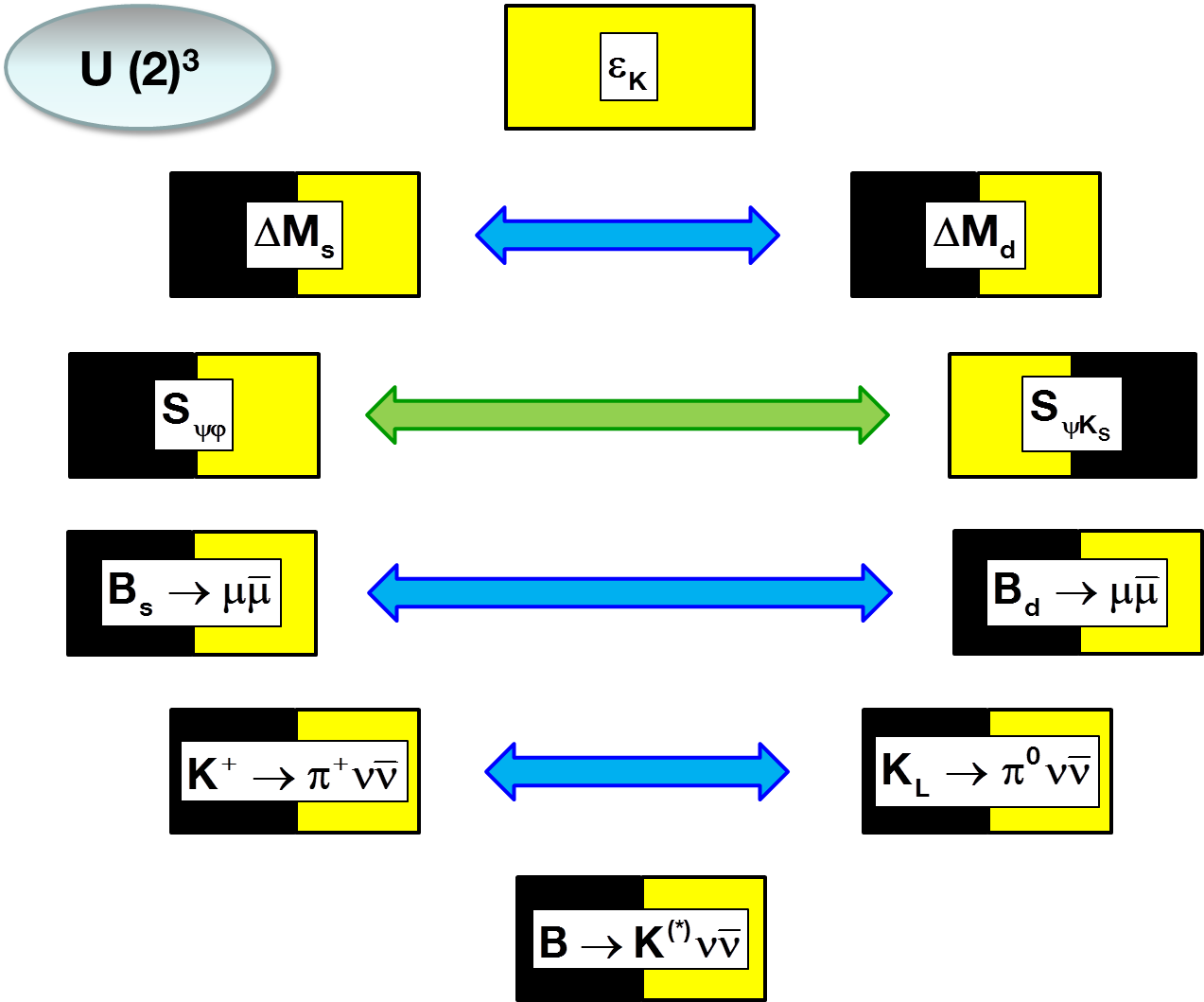}
\caption{\it DNA-chart of MFV  models (left) and of  $U(2)^3$ models (right). Yellow means   \colorbox{yellow}{enhancement}, black means
\colorbox{black}{\textcolor{white}{\bf suppression}} and white means \protect\framebox{no change}. Blue arrows
\textcolor{blue}{$\Leftrightarrow$}
indicate correlation and green arrows \textcolor{green}{$\Leftrightarrow$} indicate anti-correlation. From \cite{Buras:2013ooa}.}
 \label{fig:CMFVchart}~\\[-2mm]\hrule
\end{figure}

\begin{figure}[!tb]
\centering
\includegraphics[width = 0.49\textwidth]{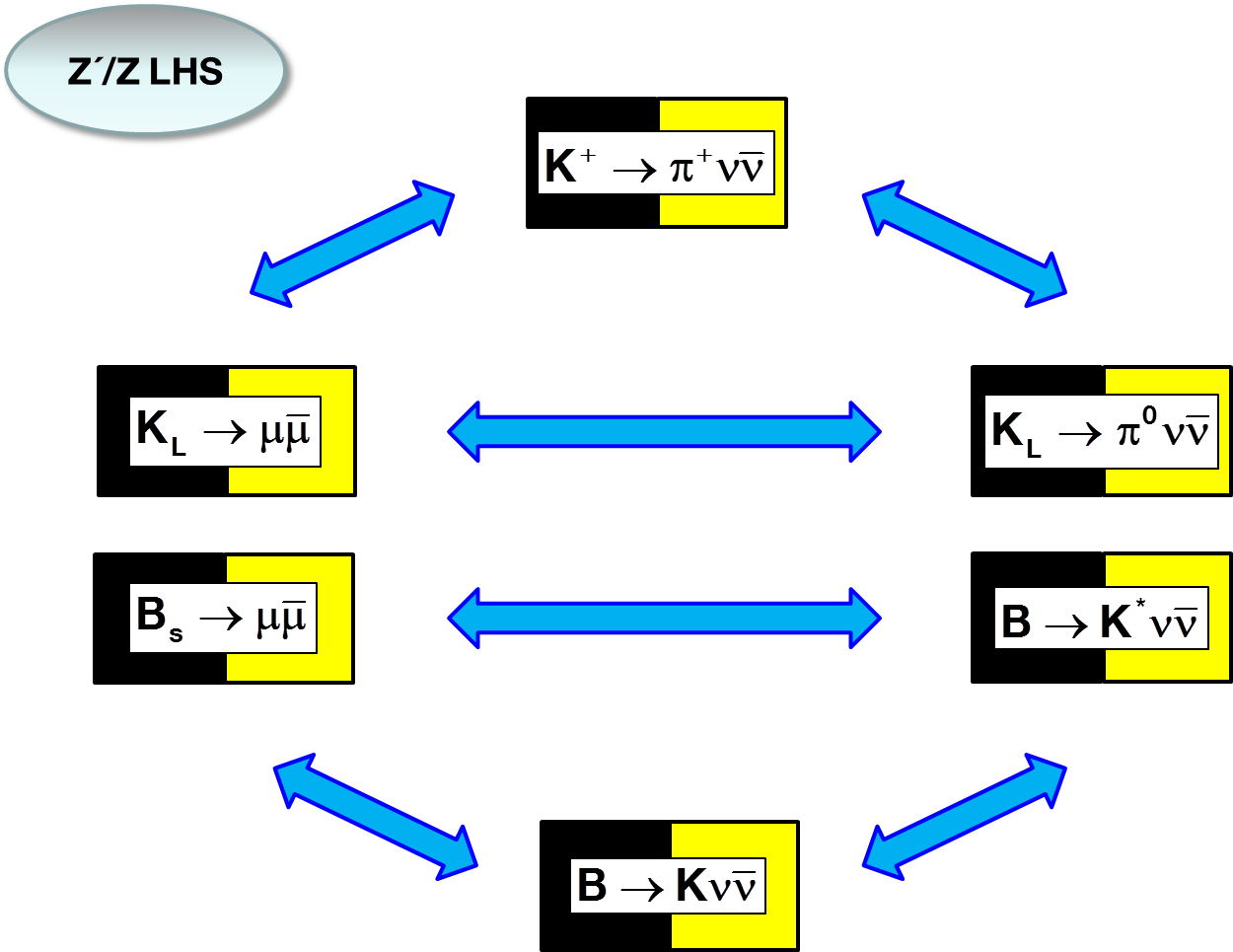}
\includegraphics[width = 0.49\textwidth]{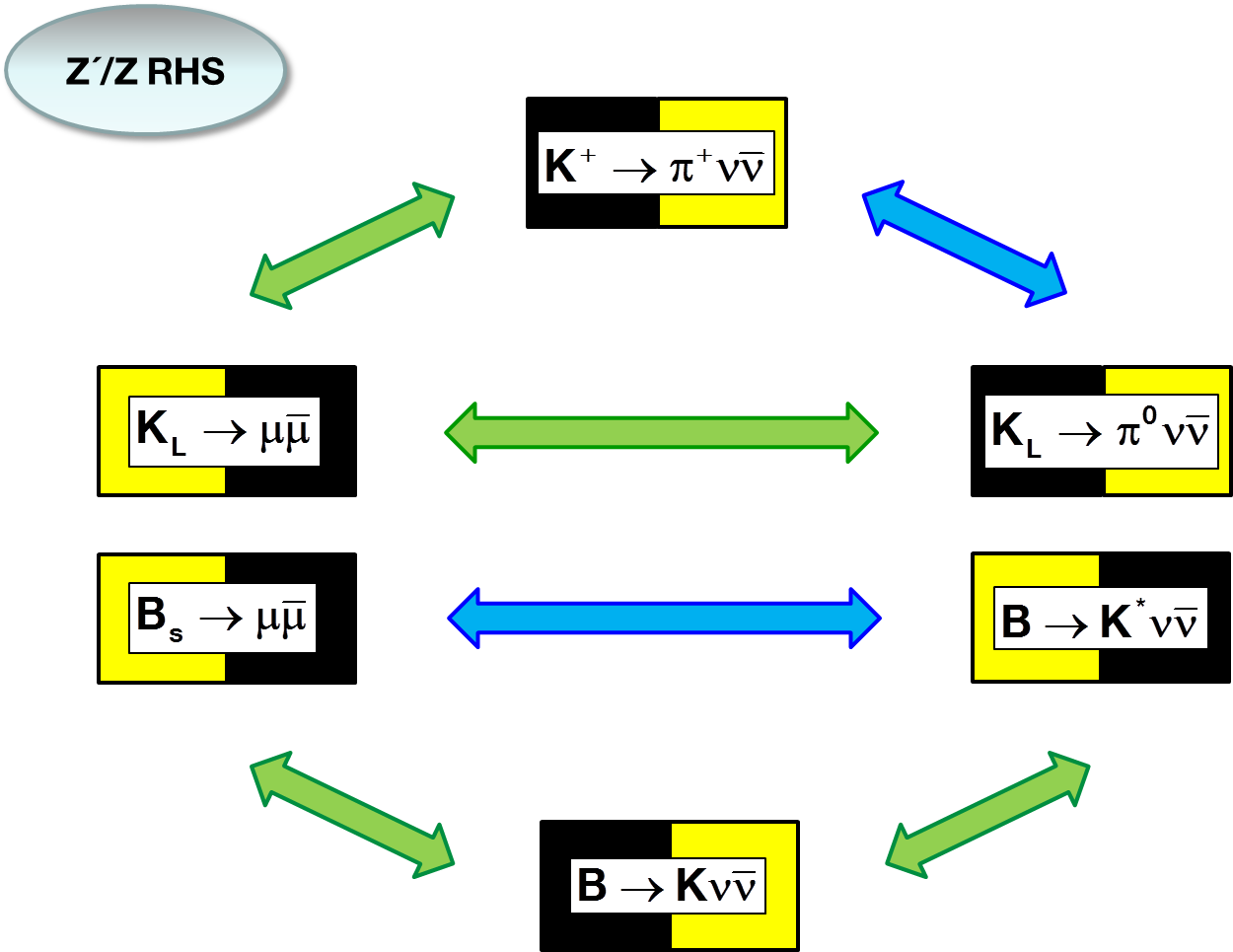}

\caption{\it DNA-charts of $Z^\prime$ models with LH and RH currents.  Yellow means   \colorbox{yellow}{enhancement}, black means
\colorbox{black}{\textcolor{white}{\bf suppression}} and white means \protect\framebox{no change}. Blue arrows
\textcolor{blue}{$\Leftrightarrow$}
indicate correlation and green arrows \textcolor{green}{$\Leftrightarrow$} indicate anti-correlation.  From \cite{Buras:2013ooa}.}
 \label{fig:ZPrimechart}~\\[-2mm]\hrule
\end{figure}

Once the CKM and hadronic parameters are sufficiently known, the search for 
NP becomes easier as now SM predictions are precise and serve as candles of 
flavour physics allowing us to see whether there is something beyond the dynamics 
we know. In this spirit, as emphasized in \cite{Buras:2013ooa}, already the pattern of signs of departures  from SM expectations in various observables and  the correlations or anti-correlations between these departures could exclude or 
support certain NP scenarios. In order to depict various possibilities in 
a transparent manner we have proposed a DNA-chart  to be applied separately 
to each NP scenario. 
In Fig.~\ref{fig:CMFVchart} we show the DNA-chart of MFV and the corresponding chart for $U(2)^3$ models.
The DNA-charts representing models with left-handed and right-handed flavour violating couplings of  $Z$ and $Z^\prime$  can be found in Fig.~\ref{fig:ZPrimechart}.

One can check that these charts summarize compactly the 
(anti-) correlations between processes of class A and B that we discussed before and also the correlations and anti-correlations within each class. In particular, the change of a correlation into an anti-correlation between two observables belonging to two different classes, when left-handed couplings 
are changed to right-handed ones, are clearly visible in these charts. We observe the following features:
\begin{itemize}
\item
Comparing   the DNA-charts of CMFV and $U(2)^3$ models 
in  Fig.~\ref{fig:CMFVchart}  we observe that the correlations 
between $K$ and $B_{s,d}$ systems are broken in the $U(2)^3$ case as the flavour symmetry is reduced from 
$U(3)^3$ down to $U(2)^3$. The anti-correlation between $S_{\psi\phi}$ and 
$S_{\psi K_S}$ is just the one shown in Fig.~\ref{fig:SvsS}.
\item
As the decays $\kpn$, $\klpn$ and $B\to K\nu\bar\nu$  belonging to class A are only sensitive
to the vector quark currents, they do not change when the couplings are changed from  left-handed to right-handed ones. On the other hand, the remaining 
three decays in   Fig.~\ref{fig:ZPrimechart} belonging to class B are sensitive to axial-vector 
couplings implying interchange of enhancements and suppressions when going from 
$L$ to $R$ and also change of correlations to anti-correlations between the 
latter three and the former three decays. Note that the correlation between 
$B_s\to\mu^+\mu^-$  and $B\to K^*\mu^+\mu^-$ does not change as both decays are  sensitive only to axial-vector coupling if in the latter case the contribution 
from the longitudinal and parallel transversity components dominate.
\item
However, it should be remarked that in order to obtain the correlations or 
anti-correlations in LHS and RHS scenarios it was assumed in the DNA charts 
presented here that the signs 
of the left-handed couplings to neutrinos and the axial-vector couplings 
to muons are the same which does not have to be the case. If they are 
opposite the correlations between the decays with neutrinos and muons in 
the final state change to anti-correlations and vice versa. 
\item
On the other hand, due to $SU(2)_L$ symmetry the left-handed $Z^\prime$
 couplings to muons and neutrinos are equal and this implies the relation
\be\label{SU2}
\Delta_{L}^{\nu\bar\nu}(Z')=\frac{\Delta_V^{\mu\bar\mu}(Z')-\Delta_A^{\mu\bar\mu}(Z')}{2}. 
\ee
Therefore, once two of these couplings are determined, the third follows uniquely without the freedom mentioned in the previous item.
\item
In the context of the DNA-charts in  Fig.~\ref{fig:ZPrimechart}, the correlations involving $\klpn$ apply only if NP contributions carry some CP-phases. If this is not the case the branching ratio for $\klpn$ will remain unchanged relative to the SM one.
\end{itemize}

If in the case of tree-level $Z^\prime$ and $Z$ exchanges 
both LH and RH quark couplings are present and are equal to each 
other (LRS scenario) or differ by sign (ALRS scenario), then one finds 
\cite{Buras:2012jb}
\begin{itemize}
\item
In LRS NP contributions to $B_{s,d}\to\mu^+\mu^-$ vanish, but they are present 
in  $\klpn$, 
$\kpn$, $B_d\to K\mu^+\mu^-$ and $B\to K\nu\bar\nu$.
\item
In ALRS NP contributions to $B_{s,d}\to\mu^+\mu^-$ are non-vanishing. 
 On the other hand 
they are absent in the case of  $\klpn$, $\kpn$, $B_d\to K\mu^+\mu^-$ and 
 $B\to K\nu\bar\nu$.
\item
In  $B_d\to K^*\mu^+\mu^-$ and  $B\to K^*\nu\bar\nu$ this rule is more complicated as already stated above, but generally the LH and RH contributions interfere
destructively in LRS and constructively in ALRS. The details depend on form factors.
\end{itemize}

\section{Adressing Anomalies in $b\to s \ell^+\ell^-$ Transitions}
\subsection{General Discussion}
We will next address the first anomalies listed in the overture. 

In addition to anomalies in angular observables in $B_d\to K^*\mu^+\mu^-$ 
found by LHCb at low $q^2$ also some departures from the SM at large $q^2$ 
in the ratios
\be
 \mathcal R_{K\mu\mu} = \frac{\mathcal{B}(B^+\to K^+\mu^+\mu^-)^{[15,22]}}{\mathcal{B}(B^+\to K^+\mu^+\mu^-)^{[15,22]}_\text{SM}},\qquad
 \mathcal R_{K^*\mu\mu} = \frac{\mathcal{B}(B^0\to K^{*0}\mu^+\mu^-)^{[15,19]}}{\mathcal{B}(B^0\to K^{*0}\mu^+\mu^-)^{[15,19]}_\text{SM}}\label{R2}
\ee
are found. Here, the superscripts refer to the range in $q^2$ in GeV${}^2$. 
Both ratios are below unity as we will see in the plots below. Considering 
large $q^2$ allows to suppress dipole operator contributions and make the 
correlations with $B_s\to \mu^+\mu^-$ and $b\to s \nu\bar\nu$ transitions more transparent.

To this end we define the ratios 
\be
 \mathcal R_{\mu\mu}= \frac{\mathcal{B}(B_s\to\mu^+\mu^-)}{\mathcal{B}(B_s\to\mu^+\mu^-)_\text{SM}}\label{R1}\,,
\ee
\be
 \mathcal R_{K} = \frac{\mathcal{B}(B\to K\nu\bar\nu)}{\mathcal{B}((B\to K\nu\bar\nu)_\text{SM}},\qquad
 \mathcal R_{K^*} = \frac{\mathcal{B}(B\to K^*\nu\bar\nu)}{\mathcal{B}((B\to K^*\nu\bar\nu)_\text{SM}}
  \,.
\ee

Interestingly the present data can be summarized by 
\be\label{ANOM1}
 \mathcal R_{\mu\mu}\approx 0.80, \qquad  \mathcal R_{K \mu\mu}\approx 0.90, \qquad  \mathcal R_{K^*\mu\mu}\approx 0.75  \,.
\ee
The important point is that they all are significantly below unity.

As shown in recent papers \cite{Descotes-Genon:2014uoa,Altmannshofer:2014rta,Hiller:2014yaa,Glashow:2014iga,Altmannshofer:2015sma} these anomalies and in 
particular those at low $q^2$ can be reproduced  by 
the shifts in the Wilson coefficients $C_9$ and $C_{10}$ with 
\be\label{910}
C_9^\text{NP}\approx -C^\text{NP}_{10} \approx -0.6 \,.
\ee
The precise values depending on the paper. The solution with NP being 
present only in $C_9$ is even favoured, but much harder to explain in the context of  existing models. We refer to \cite{Altmannshofer:2015sma} for tables 
with various solutions.

Previous discussions of this actual topic can be found in
\cite{Descotes-Genon:2013wba,Altmannshofer:2013foa,Beaujean:2012uj,Bobeth:2012vn,Buras:2013qja,Beaujean:2013soa,Datta:2013kja,Ghosh:2014awa,Hurth:2014vma}.
It should be emphasized that these analyses are subject 
to theoretical uncertainties, which have been discussed at length in 
\cite{Khodjamirian:2010vf,Beylich:2011aq,Matias:2012qz,Jager:2012uw,Descotes-Genon:2013wba,Hambrock:2013zya,Horgan:2013pva,Horgan:2013hoa,Jager:2014rwa} and it remains to be seen whether the observed anomalies are only 
result of statistical fluctuations and/or underestimated error uncertainties.

As far as theoretical uncertainties are concerned, much cleaner is the 
ratio
\be
 \mathcal R^{\mu e}_K= \frac{\mathcal{B}(B^+\to K^+\mu^+\mu^-)^{[1,6]}}{\mathcal{B}(B^+\to K^+ e^+ e^-)^{[1,6]}}=0.745^{+0.090}_{-0.074}(\text{stat})\pm 0.036(\text{syst})
\label{RLF}\,,
\ee
where the quoted value is the one from LHCb. It is by $2.6\sigma$ lower than 
its SM value: $1 +\ord(10^{-4})$ and is an intriguing signal of the breakdown 
of lepton flavour universality.

There is some consensus in the community that  all these anomalies can 
be most naturally reproduced with the help of tree-level $Z^\prime$ contribution with $Z^\prime$ having only 
left-handed flavour violating couplings to quarks \cite{Altmannshofer:2014cfa,Buras:2013qja,Buras:2014fpa,Glashow:2014iga,Bhattacharya:2014wla,Crivellin:2015mga,Altmannshofer:2014rta,Crivellin:2015lwa,Crivellin:2015era} \footnote{The $Z'$ in 331 models has  lepton flavour universal couplings~\cite{Gauld:2013qba,Gauld:2013qja,Buras:2013dea,Buras:2014yna} but can still  solve the $B\to K^*\mu^+\mu^-$ anomaly and improve on 
$B_s\to\mu^+\mu^-$. See below.} 
 or leptoquark exchanges \cite{Hiller:2014yaa,Buras:2014fpa,Gripaios:2014tna,Sahoo:2015wya,Varzielas:2015iva}, even if there exist other 
explanations \cite{Biswas:2014gga,Niehoff:2015bfa}.  As $C_9$ and $C_{10}$ 
involve vector and axial vector couplings to muons, respectively,  the $SU(2)$ relation 
in (\ref{SU2}) implies then the values of neutrino couplings. As the $Z^\prime$
couplings to quarks entering the ratios $R_i$ listed above are the same,
the present anomalies imply then uniquely  \cite{Buras:2014fpa}
\be\label{ANOM2}
 \mathcal R_{K}> 1, \qquad  \mathcal R_{K^*}> 1 \qquad (Z^\prime).
\ee
A similar exercise in the case 
of SM $Z$ boson, where the leptonic couplings are known, implies  \cite{Buras:2014fpa}
\be\label{ANOM3}
 \mathcal R_{K}<1, \qquad  \mathcal R_{K^*}< 1 \qquad (Z).
\ee
In this manner one could distinguish between these two scenarios. But in 
fact a closer look shows that $Z$ with FCNCs cannot explain these anomalies.

It is instructive to see how this different behaviour arises. In the presence 
of the dominance of left-handed flavour violating quark currents, both in the 
SM and beyond it, only one operator with $(V-A)\otimes (V-A)$ structure contributes and its Wilson coefficient is 
\be
C_L=C_L^\text{SM}+C_L^\text{NP},\qquad C_L^\text{SM}< 0\,.
\ee
Now the products of $Z^\prime$ left-handed flavour violating quark couplings  and of the muon couplings $\Delta_V^{\mu\bar\mu}(Z')$ and $\Delta_A^{\mu\bar\mu}(Z')$ build the 
Wilson coefficients $C_9^\text{NP}$ and $C_{10}^\text{NP}$, respectively.
The invariance of NP under the SM gauge group implies then, as seen in (\ref{SU2}),
\be\label{Zp}
C_L^\text{NP}= \frac{C_9^\text{NP}-C_{10}^\text{NP}}{2}\approx C_9^\text{NP},  \qquad (Z^\prime)
\ee
with the second relation following from the $b\to s \mu^+\mu^-$ data (\ref{910}). This relation can be satisfied in $Z^\prime$ models with left-handed couplings to both quarks and leptons.  With $ C_L^\text{SM}< 0$ and $C_9^\text{NP}<0$ one obtains automatically the results in (\ref{ANOM2}).

On the other hand, in the $Z$ case one has
\be\label{ZANOM}
C_L^\text{NP}= C_{10}^\text{NP}= - 13.3~ C_9^\text{NP},  \qquad (Z),
\ee
which follows not from the data, but from the known $Z$ couplings to muons. 
With $ C_L^\text{SM}< 0$ and $C_9^\text{NP}<0$ one obtains the result in (\ref{ANOM3}). But (\ref{ZANOM}) strongly disagrees with  (\ref{910}) and the explanation of $B\to K^*\mu^+\mu^-$ anomalies would imply very strong suppression of 
$\mathcal{B}(B_s\to\mu^+\mu⁻)$ relative to the SM which disagrees with the data.
On the other hand, the agreement with the data on $\mathcal{B}(B_s\to\mu^+\mu⁻)$
would allow only very small value of $C_9^\text{NP}$. It should be emphasized 
that this is also the problem of $Z$-penguins beyond the SM.

If  NP is only present in $C_9$  then $C_{10}^\text{NP}=0$ and 
 then $Z$ does not contribute to $\mathcal R_{K}$ and $\mathcal R_{K^*}$
but again (\ref{ANOM2}) is valid.

This discussion shows  that a $Z^\prime$ is the favourite scenario for the 
explanation of the $B\to K^*\mu^+\mu^-$  anomalies between these two scenarios. The problem of $Z$ can 
be also traced back to the smallness of the vector coupling of $Z$ to muons. 
Also $\mathcal R^{\mu e}_K\not=1$ being a signal of violation of lepton universality can  easily be arranged in the case of a $Z^\prime$ but not  in the case of $Z$ because of the LEP data.

Now, generally the presence of $Z^\prime$ with flavour violating couplings generates through $Z-Z^\prime$ mixing such couplings for $Z$ so that both $Z^\prime$ 
and $Z$ contribute. In this case, one finds in place of (\ref{Zp})
\be
C_L^\text{NP}= \frac{C_9^\text{NP}-C_{10}^\text{NP}}{2}+ 3\frac{C_Z}{2}, \qquad (Z^\prime,Z), 
\ee
where $C_Z$ depends on  $Z-Z^\prime$ mixing. As this mixing is model dependent nothing concrete can be concluded without having a specific model. This shows 
the superiority of the top-down approach over the bottom-up approach. The 
case of 331 models, which we will summarize now, illustrates this in a transparent manner.

\subsection{331 Models}
A concrete example for $Z^\prime$ tree-level FCNC is a model based on
the gauge group   $SU(3)_C\times SU(3)_L\times U(1)_X$, the so-called 331 model, originally developed in \cite{Pisano:1991ee,Frampton:1992wt}. There are different versions of the 331 model characterized by a parameter $\beta$ that
determines the particle content. The value of $\beta$ specifies also the leptonic couplings of $Z^\prime$ so that this model is much more predictive than 
general $Z^\prime$ models. 
In three detailed papers \cite{Buras:2012dp,Buras:2013dea,Buras:2014yna} we have analyzed various aspects of flavour violation in these models finding numerous 
correlations between various observables\footnote{There are other analyses of flavour physics which are referred to in \cite{Buras:2014yna} and in  \cite{Martinez:2014lta}.}. I will summarize here only the last of these analyzes as this paper is more complete than the previous two. It 
includes the effects of $Z-Z^\prime$ mixing which were neglected previously in 
the literature and also in our first two  papers. This turns out to be justified 
for $\Delta F=2$ processes and $B\to K^*\mu^+\mu^-$ but for certain values of $\beta$ the contributions
from the induced FCNCs mediated by $Z$ to  cannot be neglected in other processes. A new aspect of 
this paper is the correlation of flavour violating effects with the electroweak 
precision observables which allows to select the favourite values of $\beta$ 
and further selection will be possible when the flavour data improve.
Moreover, \cite{Buras:2014yna} investigates the dependence on the fermion 
representations, presenting the results for two cases. After constraints from 
$\Delta F=2$ transitions have been imposed, in addition to $\beta$ 
and $M_{Z^\prime}$, the only new parameter is  $\tan\bar\beta$ that together 
with $\beta$ describes the $Z-Z^\prime$ mixing. 

This analysis shows very clearly the superiority of the top-down approach over 
the bottom-up approach. The correlations between various flavour observables as functions of the size of $Z-Z^\prime$ mixing and the correlations with electroweak observables are totally beyond the bottom-up approach and such correlations 
will be vital in the flavour precision era. The interested reader is invited to 
look at numerous plots in \cite{Buras:2014yna} so that she (he) can better appreciate the statements just made. In what follows I will just summarize the 
most interesting results of this study.

As far as flavour physics is concerned our main findings are as follows:
\begin{itemize}
\item
NP contributions to $\Delta F=2$ transitions and decays like $B\to K^*\ell^+\ell^-$ are governed by $Z^\prime$ tree-level exchanges. 
\item 
On the other hand, for $B_{s,d}\to \mu^+\mu^-$ decays $Z$ contributions can 
be important. We find that for $\tan\bar\beta=5.0$ these contributions 
interfere constructively with $Z^\prime$ contributions enhancing NP 
 effects, while for low $\tan\bar\beta=0.2$ $Z$ contributions 
practically cancel the ones from $Z^\prime$. 
\item
Similarly $Z$ boson tree-level contributions to $B_{s,d}$ and $K$ decays 
with neutrinos in the final state can be relevant, but in this case 
the $\tan\bar\beta$ dependence is opposite to the one found for 
 $B_{s,d}\to \mu^+\mu^-$.  We find that for $\tan\bar\beta=5.0$ 
these contributions practically cancel the ones from $Z^\prime$ but
for low $\tan\bar\beta=0.2$ $Z$ contributions 
interfere constructively with $Z^\prime$ contributions enhancing NP 
 effects. 
\item
As a result of this opposite dependence on $\tan \bar \beta$ the correlations between decays with muons and neutrinos in the final state exhibit significant 
dependence on $\tan\bar\beta$ and can serve to determine this parameter in 
the future. 
\item
Our analysis of $\epe$ is to our knowledge the first one in 331 models. 
Including both $Z^\prime$ and $Z$ contributions we find that the former 
dominate, but NP effects are not large. 
\item
We also find a strict correlation between   $\epe$ and $\mathcal{B}(\klpn)$. 
The interesting feature here, 
is the decrease of $\epe$ with  increasing $\mathcal{B}(\klpn)$
 for {\it negative} $\beta$ and its increase with 
increasing $\mathcal{B}(\klpn)$   for {\it positive} $\beta$. 
\item
Imposing the electroweak precision constraints only seven among 24 combinations of $\beta$, $\tan\bar\beta$ and two fermion representation $F_1$ or $F_2$ provide better or equally good description of the electroweak precision data compared with  the SM. Interestingly, among these models none of them allows to simultaneously suppress the rate for $B_s\to\mu^+\mu^-$ and soften the $B_d\to K^*\mu^+\mu^-$ anomaly. But there are few models which either suppress the 
rate for $B_s\to\mu^+\mu^-$ or soften the $B_d\to K^*\mu^+\mu^-$ anomaly. Yet,
none of these models allows significant NP effects in $B$ and $K$ decays 
with neutrinos in the final state although departures by $15\%$ relative 
to the SM prediction for the rate of $\klpn$ are still possible.
\end{itemize}

If the $B_d\to K^*\mu^+\mu^-$ anomalies will remain, then three 331 models 
among 24 considered in  \cite{Buras:2014yna} will be favoured. These are 
M3 with $\beta=-1/\sqrt{3}$, $\tan\bar\beta=1$ and fermion representation F1, 
M14 with $\beta=1/\sqrt{3}$, $\tan\bar\beta=5$ and fermion representation F2 and  finally M16 with $\beta=2/\sqrt{3}$, $\tan\bar\beta=5$ and fermion representation F2. However, none of these models can describe breakdown of lepton flavour universality in $B_d\to K\ell^+\ell^-$ observed by the LHCb.

\section{The Power of $B\to K^{(*)}\nu\bar\nu$, $\kpn$ and $\klpn$  Decays}\label{sec:8}
\subsection{$B\to K^{(*)}\nu\bar\nu$}
We have just seen that these decays, when measured, could allow to distinguish 
between various explanations of the present anomalies in $b\to s \mu^+\mu^-$ 
transitions. But it should be stressed that these decays are of interest on 
its own as they are theoretically  cleaner than $B\to K^{(*)}\mu^+\mu^-$ and 
allow good tests of the presence of right-handed currents and in general of NP.

Both decays should be measured 
at Belle II. The most recent estimate of their branching ratios within the SM  \cite{Buras:2014fpa} reads:
\be
\mathcal{B}(B^+\to K^+\nu\bar\nu) = \left[\frac{\vcb}{0.0409}\right]^2(4.0 \pm 0.4) \times 10^{-6}, 
\ee
\be
\mathcal{B}(B^0\to K^{* 0}\nu\bar\nu)  = \left[\frac{\vcb}{0.0409}\right]^2 (9.2\pm 0.9) \times 10^{-6},
\ee
where the errors in the parentheses are fully dominated by form factor uncertainties. We expect that when these two branching ratios will be measured, these 
uncertainties will be further decreased and $\vcb$ will be precisely known so 
that a very good test of the SM will be possible.

But in the context of such tests one should still take care of non-perturbative tree level contributions from $B^+\to
\tau^+\nu$ to $B^+\to K^+\nu\bar\nu$ and $B^+\to K^{*+}\nu\bar\nu$ at the 
level of roughly $(5-10)\%$  which have  been pointed out \cite{Kamenik:2009kc}.
This should be possible in the future ones the data on  $B^+\to
\tau^+\nu$ will be precise. The SM results quoted above 
refer only to the short-distance contributions.

An extensive analysis of these decays model-independently and in various extensions of the SM has been performed in  \cite{Buras:2014fpa}. In addition 
to the correlations between various ratios $R_i$ discussed for $Z^\prime$ 
models, shown in figure 5 of that paper, of particular interest are the correlations between $\mathcal R_{K}$ and $\mathcal R_{K^*}$ in various scalar and vector leptoquarks models with leptoquarks carrying different quantum numbers.  Figure 10 in \cite{Buras:2014fpa} shows that precise measurements of  $\mathcal R_{K}$ and $\mathcal R_{K^*}$ 
could distinguish between various leptoquark models. 

Also  
figures 1, 2, 3  and 11 in \cite{Buras:2014fpa}
demonstrate very clearly that we will have 
a lot of fun when the experimental data on $B\to K^{(*)}\mu^+\mu^-$ and $B\to K^{(*)}\nu\bar\nu$ will be known and the theoretical issues in $B\to K^{(*)}\mu^+\mu^-$ are clarified. But we will have even more fun when the branching ratios on 
$\kpn$ and $\klpn$ will be measured.
\subsection{ $\kpn$ and $\klpn$ in the SM}
 In view of the recent start of the NA62 experiment at CERN that is expected to measure the $\kpn$ branching ratio with the precision of $10\%$  \cite{Rinella:2014wfa,Romano:2014xda}, we have recently summarized the present  status of this decay  within the SM and of $\klpn$ which should be measured by KOTO experiment around 2020 at J-PARC \cite{Komatsubara:2012pn,Shiomi:2014sfa}. As the perturbative QCD \cite{Buchalla:1993bv,Misiak:1999yg,Buchalla:1998ba,Buras:2005gr,Buras:2006gb,Gorbahn:2004my} and electroweak corrections \cite{Brod:2008ss,Brod:2010hi,Buchalla:1997kz}
in both decays are fully under control, the present uncertainties in the branching ratios originate within the SM dominantly from $\vcb$, $\vub$ and $\gamma$ when extracted from tree-level decays. Unfortunately  the clarification of the discrepancies between inclusive and exclusive determinations of $\vcb$ and $\vub$ from tree-level decays are likely to be resolved only at the time of  the Belle II experiment at  SuperKEKB  at the end of this decade. Therefore, 
we investigated  whether in the coming years higher precision on both 
branching ratios can be obtained by eliminating $\vcb$, $\vub$ and $\gamma$ with the help of other observables that are already precisely measured. In this context $\varepsilon_K$ and  $\Delta M_{s,d}$, accompanied by  recent and future 
 progress in QCD lattice calculations, as well as the improved measurements of mixing induced CP asymmetries $S_{\psi K_S}$ and $S_{\psi\phi}$ at the LHC  will play prominent roles. 

We find \cite{Buras:2015qea} 
\be
\mathcal{B}(\kpn)= \left(9.1\pm 0.7\right) \times 10^{-11}, \qquad 
\mathcal{B}(\klpn)= \left(3.0\pm 0.3 \right) \times 10^{-11}
\ee
which is roughly by a factor of two more precise than using present tree-level values of $\vcb$, $\vub$ and $\gamma$ in (\ref{average}) and (\ref{gamma}).

 We also find
\be
\vcb= (42.4\pm 1.2)\times 10^{-3}, \qquad \vub= (3.61\pm 0.14)\times 10^{-3}, 
\qquad \gamma=(69.5\pm 5.0)^\circ
\ee
 in this manner. The large value of $\vcb$ is fully consistent with 
its inclusive determinations and is required by the data on $\varepsilon_K$. 
We note that with the reduced error on the $\Delta B=2$ paramete $\xi$, promised in \cite{Bouchard:2014eea}, the error on $\gamma$ will decrease down to $2.3^\circ$.
 
The indirect fits done by UTfit \cite{Bona:2009cj} and CKMfitter \cite{Charles:2011va} that are summarized in \cite{Ricciardi:2014aya} yield
\begin{align}
 \text{UTfit: }\quad&\vub =(3.63\pm0.12)\times 10^{-3}, \qquad  \vcb=(41.7\pm0.6)\times 10^{-3}\,,\label{equ:UTfit}\\
\text{CKMfitter: }\quad&\vub =\left(3.57^{+0.41}_{-0.31}\right)\times 10^{-3}, \qquad \,\,\,\,\vcb=\left(41.4^{+1.4}_{-1.8}\right)\times 10^{-3}\,\label{equ:CKMfitter}\,,
\end{align}
in good agreement with our results. We note however, that these two groups 
included in their analyses the information from tree level decays, whereas 
we decided not to use it because of discrepancies  between inclusive and 
exclusive determinations of $\vub$ and $\vcb$. We also left out tree level 
determination of $\gamma$ from this fit.

\begin{figure}[t]
\centering%
\includegraphics[width=0.455\textwidth]{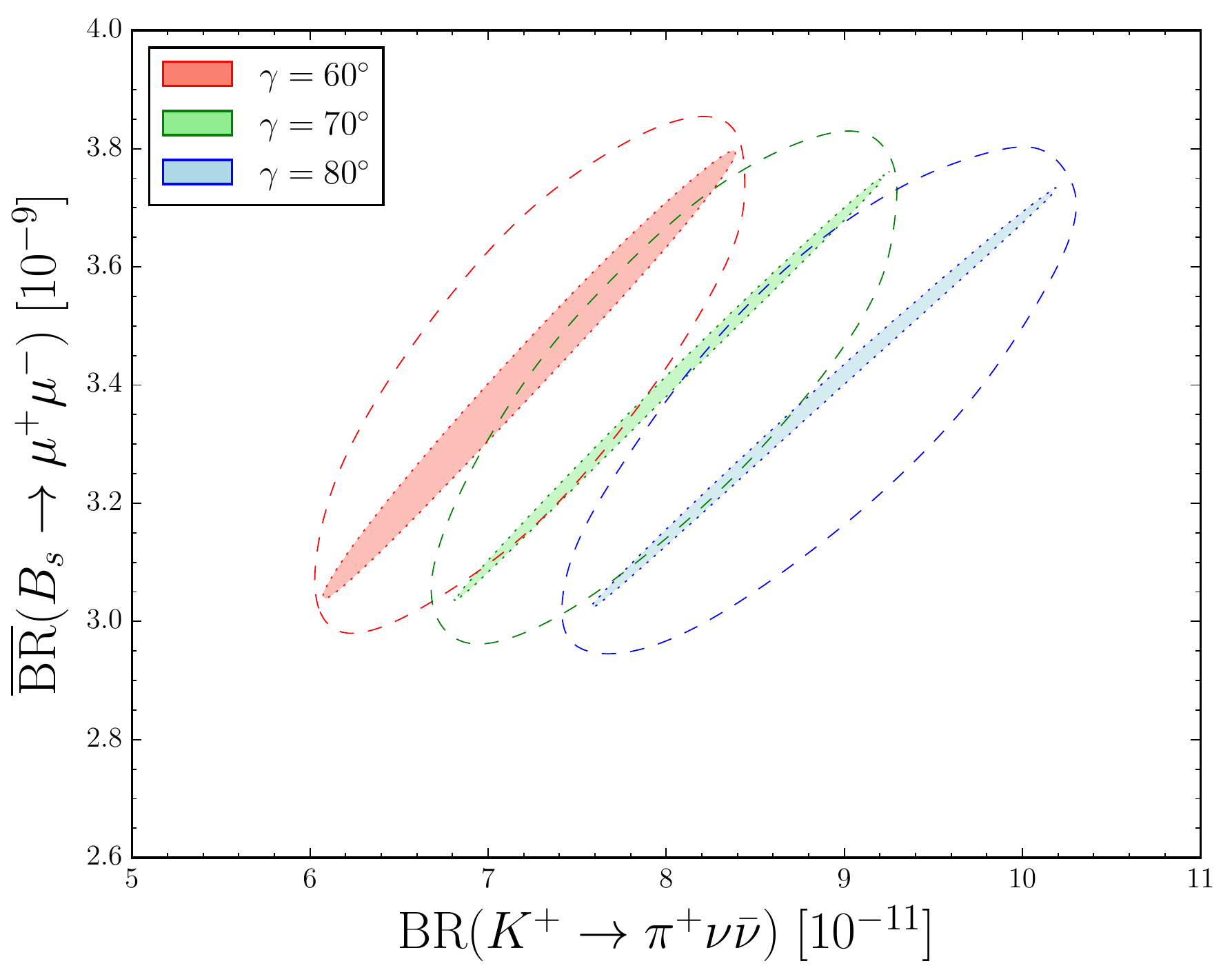}
\includegraphics[width=0.45\textwidth]{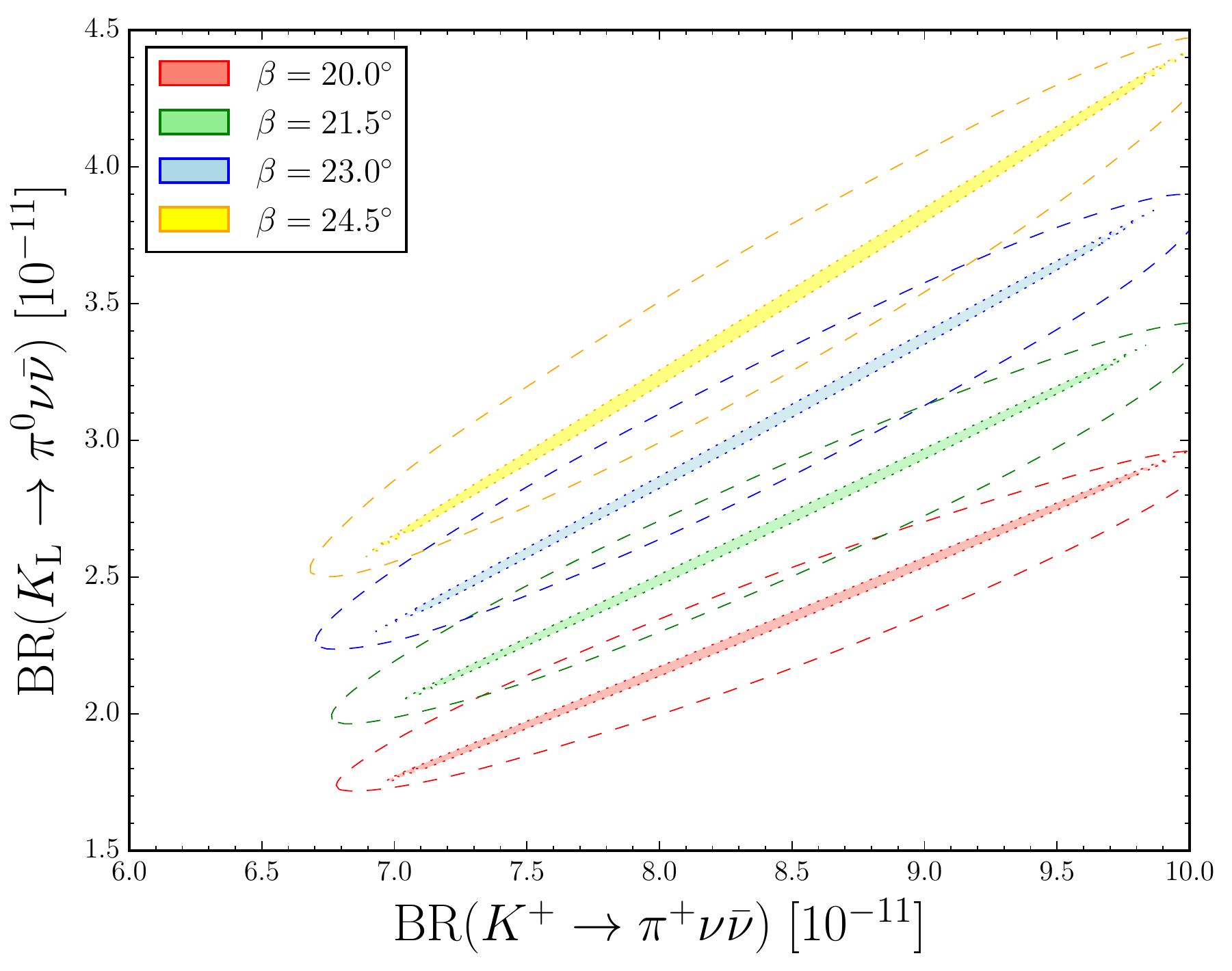}%
\caption{\it Left panel: correlation of $\overline{\mathcal{B}}(B_s\to\mu^+\mu^-)$ versus $\mathcal{B}(\kpn)$ in the SM for fixed values of $\gamma$. Right panel: correlation of ${\mathcal{B}}(\klpn)$ versus $\mathcal{B}(\kpn)$ in the SM for fixed values of $\beta$. In both plots the dashed regions correspond to a 68\% CL resulting from the uncertainties on all other inputs, while the inner filled regions result from including only the uncertainties from the remaining CKM inputs. From  \cite{Buras:2015qea}. \label{fig:fixedPlots}}
\end{figure}

In my view one of the highlights of \cite{Buras:2015qea} is the new correlation 
between $\mathcal{B}(\kpn)$, $\mathcal{B}(B_s\to\mu^+\mu^-)$ and $\gamma$ extracted from tree-level decays within the SM that is only very weakly dependent on  other CKM parameters, in particular $\vub$ and $\vcb$.
This correlation should be of interest to experimentalists from the LHCb, 
CMS and NA62, 
who in the coming years will significantly improve the measurements on 
these three quantities. We show it in the left panel of Fig.~\ref{fig:fixedPlots}. We observe that the central experimental value of  $\overline{\mathcal{B}}(B_s\to\mu^+\mu^-)$, that is below $3.0 \times 10^{-9}$, implies  $\mathcal{B}(\kpn)$ in the ballpark $7.0 \times 10^{-11}$. If confirmed, this would possibly 
make the identification of NP in latter decay easier.

There are other interesting correlations presented in  \cite{Buras:2015qea}, like the 
update of the correlation between branching ratios for $\kpn$ and $\klpn$ 
and $S_{\psi K_S}$ that is practically independent of  $\vub$ and $\vcb$ 
 \cite{Buchalla:1994tr,Buras:2001af}. We show this correlation in the right 
panel of Fig.~\ref{fig:fixedPlots}. 
Moreover, we provide a number of 
rather accurate expressions that should allow one to monitor easily the future 
experimental and theoretical progress on the observables in question.
\subsection{ $\kpn$ and $\klpn$ beyond the SM}
Beyond the SM both branching ratios can still be significantly enhanced, but the 
size of this enhancement strongly depends on NP scenario considered and the constraints in a given scenario coming from other observables. A new analysis in progress and other analyses by us, in particular in  \cite{Buras:2014sba}, allow us to draw a general picture of NP contributions to the two decays in question:
\begin{itemize}
\item
In models with MFV and $U(2)^3$ models NP effects amount to at most $20-30\%$ 
at the level of the branching ratio making the distinction from the SM difficult. The characteristic feature in these scenarios is 
the strong correlation between both branching ratios with both either enhanced 
or suppressed relative to the SM.
\item
In non-MFV models the shape of the correlation depends on whether $\varepsilon_K$ constraint is important or not \cite{Blanke:2009pq}. If it is, the shape of this correlation is 
as shown in the left-panel of Fig.~\ref{KLKP}. The upper brunch is
characteristic for MFV, while the lower one shows a non-MFV behaviour: the 
branching ratio for $\kpn$ is modified while the one for $\klpn$ remains basically 
unchanged. On the other hand, if with sufficient number of parameters  $\varepsilon_K$ constraint can be eliminated then the correlation has the shape shown in 
the right panel of  Fig.~\ref{KLKP}  \cite{Buras:2014zga}.
\item
The size of possible enhancements of $\mathcal{B}(\klpn)$ is bounded by 
$\epe$. The  size of possible enhancements of $\mathcal{B}(\kpn)$  by $K_L\to\mu^+\mu^-$ and $\epe$. This has been known already for many years 
\cite{Buras:2013ooa} but in 
view of the improved determination of electroweak contribution to $\epe$ with 
the help of lattice QCD, such studies become more quantitative. We refer to 
 \cite{Buras:2014sba}  for details.
\item
In a scenario with arbitrary flavour-violating $Z$ couplings to quarks the 
correlation of $\klpn$ with $\epe$ very significantly limits possible enhancements of  $\mathcal{B}(\klpn)$ but enhancements by a factor $3-4$ are still possible.  $\mathcal{B}(\kpn)$ can be in particular enhanced when $K_L\to\mu^-\mu^+$ 
can be eliminated. This is the case of LH and RH flavour violating $Z$ couplings  being approximately equal. Also, when only RH couplings are present. Therefore
values of $\mathcal{B}(\kpn)$ in the ballpark of  $(15-20)10^{-11}$ are still 
in principle possible.
\item
Larger enhancements are still possible in the $Z^\prime$ scenarios with arbitrary flavour-violating couplings. Here, in contrast to $Z$ models the diagonal quark couplings, required for $\epe$ are only known in concrete models. While in 
the 331 model, discussed previously,   $Z^\prime$ effects have been found to be small
 \cite{Buras:2014yna}, in general $Z^\prime$ scenarios a strict correlation between $\epe$ and both rare decays does not really exist and larger effects are 
still  allowed.
\end{itemize}

\section{Can we Reach the Zeptouniverse with Rare $K$ and $B_{s,d}$ Decays?}
We will finally address the central issue of our paper, the reach of flavour physics 
in testing very short distances scales. This issue becomes relevant independently of wheather LHC will discover new particles or not. The point is that LHC
 will directly probe only distance scales down to $10^{-19}~{\rm m}$, corresponding to energy scales at the level of a few $\tev$. This is of course impressive 
but one would like to know what happens at even shorter distance scales. 
In order to reach even higher resolution before the advent of future high-energy colliders, it is necessary to consider indirect probes of NP, a prime example being $\Delta F=2$ neutral meson mixing processes, which are sensitive to much shorter distance scales.

In fact in the framework of effective theories,  the analyses in 
\cite{Bona:2007vi,Isidori:2010kg,Charles:2013aka},
 which dealt dominantly with $\Delta F=2$ observables, have 
 shown that in the presence of left-right operators and $\ord(1)$ couplings one could be in 
principle sensitive to scales as high as $10^4\tev$, or even higher energy 
scales. Unfortunately, as pointed out in  \cite{Buras:2014zga}, $\Delta F=2$ observables 
alone will not really give 
us significant information about the particular nature of this NP. This is 
related to the symmetric structure of the formulae like the one in (\ref{LRstructure}): one cannot distinguish between left and right by using only $\Delta F=2$ transitions. On the other hand, our  DNA charts demonstrate that $\Delta F=1$ processes, in particular rare $K$ and 
$B_{s,d}$ decays, can help us in this matter through various correlations between observables that depend on whether left-handed or right-handed couplings are 
involved.

But as left-right operators involving 
four quarks are not the driving force in these decays, which generally contain 
operators built out of one quark current and one lepton current, it is not 
evident that rare $K$ and $B_{s,d}$ decays can help us in reaching the Zeptouniverse even in the flavour precision era. This interesting question has been 
addressed in \cite{Buras:2014zga} and I would like to summarize the results of this study.

Certainly the answer to this question depends on the size of NP, its nature and in particular on the available precision of experiments and 
of the SM predictions for flavour observables. The latter precision depends on the extraction of CKM parameters from the data and on the
theoretical uncertainties. Both are expected to be reduced in this decade 
down to 1\,--\,2\%, which should allow NP to be identified even if it contributed only at the level of 20\,--\,30\% to the branching ratios. 

In order to find the maximal resolution one has to decide what is the largest
coupling of NP particles to SM particles still consistent with perturbativity.
A coupling of at most $3.0$ at the high scale seems to be a reasonable choice. The results for 
resolutions quoted below depend on this number linearly. For smaller couplings 
 the resolutions are worse.

\begin{figure}[!tb]
\centering%
\includegraphics[width = 0.45\textwidth]{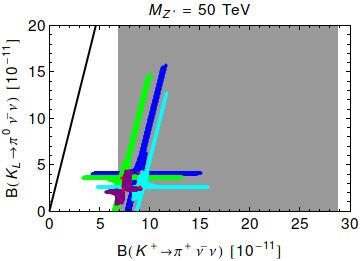}
\includegraphics[width = 0.45\textwidth]{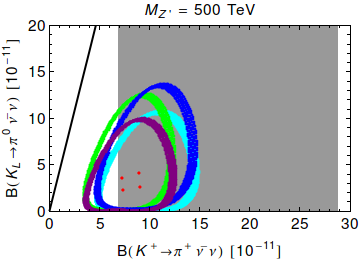}
\caption{\it $\mathcal{B}(\klpn)$ versus
$\mathcal{B}(\kpn)$ for $M_{Z^\prime} = 50~{\rm TeV}$ in the LHS (left) and for 
$M_{Z^\prime} = 500~{\rm TeV}$ in L+R scenario. The colours distinguish between different CKM input which also implies the four red points corresponding to
the SM central values for these four CKM scenarios, respectively. The black line corresponds to the Grossman-Nir bound \cite{Grossman:1997sk}. The gray region shows the
experimental range of $\mathcal{B}(\kpn))_\text{exp}=(17.3^{+11.5}_{-10.5})\times 10^{-11}$. From \cite{Buras:2014zga}.}\label{KLKP}
\end{figure}

\begin{figure}
\centering%
\raisebox{1cm}{\includegraphics[width=0.65\textwidth]{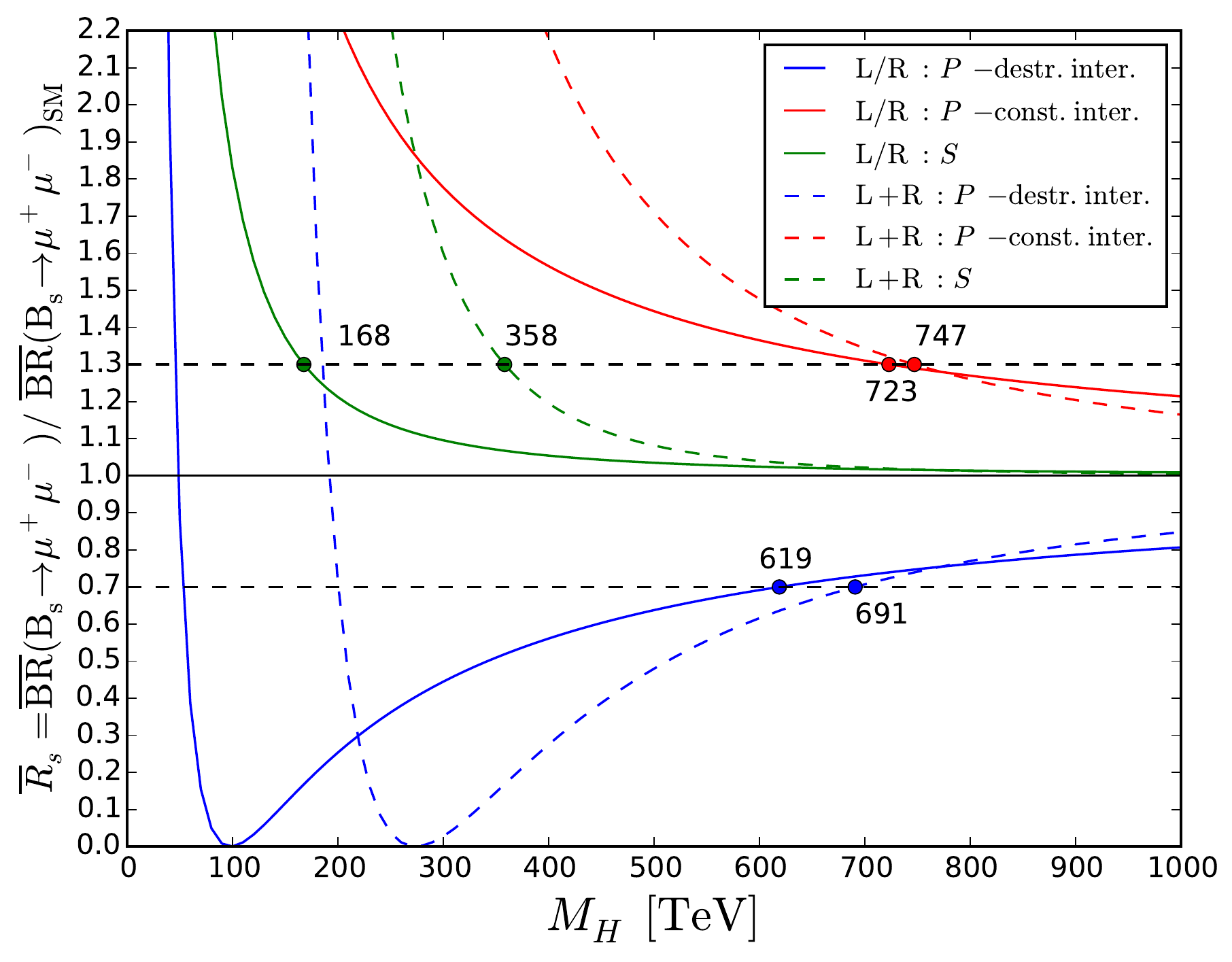}}\hfill%
\vskip-1.40cm
\caption{\it Dependence of $\overline{R}_s$ on the heavy scalar mass $M_H$, showing the pure LH 
(or RH) scenario and the combined L+R scenario (see \cite{Buras:2014zga} for details). }\label{fig:scalarMH}
\end{figure}

Answering this question first in the context of $Z^\prime$ tree-level exchanges  our main findings  are as follows:
\begin{itemize}
\item
Future precise measurements of several $\Delta F=1$ observables 
and in particular correlations between them can distinguish between LH and 
RH currents, but the maximal resolution consistent with perturbativity 
strongly depends on whether only LH   or only RH  or both LH and 
RH flavour changing $Z^\prime$ couplings to quarks are present in nature.
\item
If  $Z^\prime$ has only LH or RH couplings we can in principle reach
scales of $200\tev$ and $15\tev$ for $K$ and $B_{s,d}$, respectively.
These numbers depend on the room left for NP in $\Delta F=2$ observables, which have an important impact on the resolution available in these NP scenarios. 
In the left panel in Fig.~\ref{KLKP} we show the result of $K\to\pi\nu\bar\nu$ for $M_{Z^\prime}=50\tev$.
\item
    Smaller distance scales can only be resolved if both RH and LH couplings are present in order to cancel the NP effects in the $\Delta F=2$ observables. Simply having more free parameters one can easier satisfy $\Delta F=2$ constraints 
without relevant impact on $\Delta F=1$ transistions. But to achieve this 
some tuning of couplings is required. In particular, RH and LH couplings have to
 differ considerably from each other. This large hierarchy of couplings is dictated primarily by the ratio of hadronic matrix elements of 
LR $\Delta F=2$ operators and those for LL and RR operators (see comments 
after (\ref{LRstructure})) and by 
the room left for NP in $\Delta F=2$ processes. Future advances in the determination of CKM parameters and calculation of the relevant hadronic matrix elements  should specify this room precisely. We find that in this case 
the scales as high as $2000\tev$ can be reached with the help of 
$\kpn$ and $\klpn$ and  $160\tev$ with the help $B_{s,d}\to\mu^+\mu⁻$ when 
$Z^\prime$ is at work. In the right panel in Fig.~\ref{KLKP} we show the result of $K\to\pi\nu\bar\nu$ for $M_{Z^\prime}=500\tev$. We observe that 
the structure of this correlation is very different from the case when 
$\Delta F=2$ constraint is present. We discuss this issue below.
\item
A study of tree-level (pseudo-)scalar exchanges shows that in this case $B_{s,d}\to \mu^+\mu^-$ can probe scales up to $750\tev$, both for scenarios with purely LH or RH scalar couplings to quarks and for scenarios allowing for both 
LH and RH couplings.
For the limit of a degenerate scalar and pseudoscalar NP effects in $\Delta F=2$ observables can cancel even without imposing a tuning on the couplings. The 
outcome of this analysis is shown in Fig.~\ref{fig:scalarMH}.
\item
We have discussed models with several gauge bosons. Also in 
this case the basic strategy for being able to explore very high energy scales is to break the stringent correlation between $\Delta F=1$ and 
$\Delta F=2$ processes and to suppress NP contributions to the latter without suppressing NP contributions to rare decays. The presence of a second heavy neutral gauge boson allows us to achieve the goal with only LH or RH currents 
by applying an appropriate tuning.
\item
While the highest achievable resolution in the presence  of several gauge bosons is comparable to the case of a single  gauge boson because of the perturbativity bound, the correlations between $\Delta F=1$ observables could 
differ from the ones presented here. This would be in particular the case if 
LH and RH couplings of these bosons where of similar size. But a detailed study 
of such scenarios would require the formulation of concrete models.
\item
    If FCNCs only occur at one loop level the highest energy scales that 
can be resolved for maximal couplings are typically reduced  relative to the case of tree-level FCNCs by a factor 
of at least $3$ and $6$ for $\Delta F=1$ and $\Delta F=2$ processes, respectively.
\end{itemize}

Now comes the following difficulty, which requires further study. The observables depend generally on the ratio of the coupling and the mass of the exchanged object. Therefore, if in the future one will see some deviations from the SM, to first approximation, only this ratio will be determined and at first side without 
knowing the coupling we will not know which scales are involved. However, the 
 case of $\kpn$ and $\klpn$ decays, as seen in Fig.~\ref{KLKP}, allows to get an idea which scales are involved and this is because one of 
the decays ($\klpn$) is CP-violating. If $\varepsilon_K$ constraint is vital 
then the structure of correlation is as seen in the left panel of this figure.
But when it is eliminated, the new phase in the decays is free and a different 
correlation is seen in the right panel allowing the two branching ratios to 
take values which are not allowed if $\Delta F=2$ constraint is relevant. 
The analytic understanding of this difference has been provided in \cite{Blanke:2009pq}. Thus, if in the future  such values will be measured, they could  indeed signal that we deal here with very high scales. Moreover, this would indicate that both left-handed and right-handed currents in $\Delta F=2$ processes  at work, thereby allowing $\Delta F=2$ constraints to be satisfied.

 But when two CP conserving quantities are involved, the correlations  between observables in $B_{s,d}$ decays turn out to have similar shape independently whether both left-handed and right-handed couplings are at work. Therefore, in this 
case it is more difficult to find out which energy scales are involved. Still we have presented 
a simple idea for a rough indirect determination of  $M_{Z^\prime}$ by means of the next linear $e^+e^-$ or $\mu^+\mu^-$ collider and precision flavour data. It uses 
the fact that the LR operators present in $\Delta F=2$ transitions have 
large anomalous dimensions so that $M_{Z^\prime}$ can be determined through 
renormalisation group effects provided it is well above the LHC scales. 

In summary we have demonstrated in \cite{Buras:2014zga} that NP with a particular pattern of dynamics could be investigated in principle through rare $K$ and $B_{s,d}$ decays even if the scale of this NP would belong to the Zeptouniverse. As 
expected from other studies it is in principle easier to reach
 the Zeptouniverse 
with the help of rare $K$ decays than $B_{s,d}$ decays. However, this assumes 
the same maximal couplings in these three systems and this could be not the 
case. Moreover, in the presence of tree-level pseudoscalar exchanges very 
short distance scales can be probed by $B_{s,d}\to\mu^+\mu^-$ decays.

We should emphasise that although  the main goal in \cite{Buras:2014zga}
  was to reach the 
highest energy scales with the help of rare decays, it will  of course be exciting 
to explore any scale  of NP above the LHC scales in this decade. Moreover, we 
still hope that high energy proton-proton collisions at the LHC will exhibit at least some footprints of new particles and forces. This would greatly facilitate flavour analyses as the one just presented.

Moreover, it should not be forgotten that
in principle much higher energy scales or much better resolution could be achieved in the future with the help of charged lepton flavour 
violating decays such as $\mu\to e \gamma$, $\mu\to 3e$ and  $\tau\to 3\mu$, $\mu\to e$ 
conversion in nuclei, and electric dipole moments 
\cite{Hewett:2012ns,Engel:2013lsa,McKeen:2013dma,Moroi:2013sfa,Moroi:2013vya,Eliaz:2013aaa,Kronfeld:2013uoa,deGouvea:2013zba,Bernstein:2013hba,Altmannshofer:2013lfa}. But this topic is another story.

Finally, it should be emphasized that the analysis  \cite{Buras:2014zga} was 
just a first more detailed look beyond the LHC scales. In order to map out the 
scales from the LHC scales down to the Zeptouniverse and beyond it, much more work is needed. It requires a full expedition, analogous  the ones in the Himalayas.

\section{Final Remarks}
It is clear that presently available precision on the multitude of observables considered 
in our papers and briefly in this lecture is insufficient for the execution of the 
flavour program outlined by us. Yet, the coming flavour precision era, in 
which the measurements of most observables listed in Fig.~\ref{Fig:1} will 
reach much higher precision and lattice QCD calculations will be significantly 
improved, should allow us to obtain at least a rough picture of the physics 
beyond the LHC and if we are lucky even reach the Zeptouniverse. This would 
be a very important step towards the construction of a fundamental theory of 
particles and interactions.

Having this in mind I do not share the frustration of some of my colleagues 
caused by the lack of NP signals in high energy collisions at the LHC or difficulties in finding dark matter particles and axions. One should not forget that low energy processes and in particular rare processes like rare kaon decays and 
CP violation were vital in the construction of the SM well before the discovery of $W$ and $Z$ bosons, of the top quark and the Higgs. While 
I hope very much that in the coming years LHC will discover new particles, this  lecture shows that particle physics will still have much to offer if this will turn out not to be the case.

{\bf  Acknowledgements}\\
I thank all my collaborators, in particular Dario Buttazzo, Fulvia De Fazio, 
Jennifer Girrbach-Noe and Rob Knegjens, Christoph Niehoff and David Straub for exciting time we spent together 
exploring recently the short distance scales with the help of flavour violating 
processes. Finally, I would like to thank the organizers of FWNP for their hospitality and perfect organization of this memorable event.
The research presented in this report was dominantly financed and done in the context of the ERC Advanced Grant project ``FLAVOUR'' (267104).  It was also partially supported by the 
DFG cluster of excellence ``Origin and Structure of the Universe''.

\bibliographystyle{JHEP}
\bibliography{allrefs}
\end{document}